\documentclass[12pt,a4paper]{article}
\usepackage{amsmath}
\usepackage{amssymb}
\usepackage[nosort]{cite}
\usepackage{hyperref}
\usepackage{graphicx}
\usepackage{color}
\usepackage{float}
\allowdisplaybreaks

\makeatletter
\@addtoreset{equation}{section}

\makeatother

\makeatletter
\let\old@makecaption=\@makecaption
\def\@makecaption{\small\old@makecaption}
\makeatother

\makeatletter
\let\old@startsection=\@startsection
\renewcommand{\@startsection}[6]{\old@startsection{#1}{#2}{#3}{#4}{#5}{#6\mathversion{bold}}}
\makeatother

\let\oldPhi=\Phi
\let\oldPsi=\Psi
\let\oldGamma=\Gamma
\let\oldSigma=\Sigma
\renewcommand{\Phi}{\mathnormal{\oldPhi}}
\renewcommand{\Psi}{\mathnormal{\oldPsi}}
\renewcommand{\Gamma}{\mathnormal{\oldGamma}}
\renewcommand{\Sigma}{\mathnormal{\oldSigma}}

\newcommand{\hypref}[2]{\ifx\href\asklfhas #2\else\href{#1}{#2}\fi}

\newcommand{\alg}[1]{\mathfrak{#1}}

\newcommand{\Tr}{\mathop{\mathrm{Tr}}}

\newenvironment{myeqnarray}{\arraycolsep0pt\begin{eqnarray}}{\end{eqnarray}\ignorespacesafterend}
\newenvironment{myeqnarray*}{\arraycolsep0pt\begin{eqnarray*}}{\end{eqnarray*}\ignorespacesafterend}

\def\[{\begin{equation}}
\def\]{\end{equation}}
\def\<{\begin{myeqnarray}}
\def\>{\end{myeqnarray}}

\ifx\href\asklfhas\newcommand{\href}[2]{#2}\fi

\begin{document}

\thispagestyle{empty}

\vspace{1cm}

\renewcommand{\thefootnote}{\fnsymbol{footnote}}
\setcounter{footnote}{0}

\begin{center}

{\Large\textbf{\mathversion{bold}Three-Point Functions and $\alg{su}(1|1)$ Spin Chains}\par}

\vspace{1cm}

\textsc{Jo\~ao Caetano$^{a,b,c}$ and Thiago Fleury$^{d}$ }
\vspace{8mm}

\textit{$^{a}$ Perimeter Institute for Theoretical Physics, \\
Waterloo, Ontario N2L 2Y5, Canada}
\vspace{4mm}

\textit{$^{b}$ Department of Physics and Astronomy $\&$ Guelph-Waterloo Physics Institute,
University of Waterloo, Waterloo, Ontario N2L 3G1, Canada}

\vspace{4mm}
\textit{$^{c}$ Centro de F\'isica do Porto e Departamento de F\'isica e Astronomia, \\
Faculdade de Ci\^encias da Universidade do Porto, \\
Rua do Campo Alegre, 687, 4169-007 Porto, Portugal}

\vspace{4mm}

\textit{$^{d}$ 
Instituto de F\'isica Te\'orica, UNESP - Univ. Estadual Paulista, \\
ICTP South American Institute for Fundamental Research, \\
Rua Dr. Bento Teobaldo Ferraz 271, 01140-070, S\~ao Paulo, SP, Brasil} 

\vspace{8mm}

\texttt{jd.caetano.s@gmail.com}\\
\texttt{tfleury@ift.unesp.br}\par\vspace{1cm}

\textbf{Abstract}\vspace{5mm}

\begin{minipage}{13.7cm}
We compute three-point  functions of general operators in the $\alg{su}(1|1) $ sector of planar ${\cal N}=4$ SYM  in the weak coupling regime, both at tree-level and one-loop. 
Each operator is represented by a closed spin chain Bethe state characterized by a set of momenta 
parameterizing the fermionic excitations. 
At one-loop, we calculate both the two-loop Bethe eigenstates and the relevant Feynman diagrams for the three-point functions within our setup. The final expression for the structure constants is surprisingly simple and hints at a possible form factor based approach yet to be unveiled. 

\end{minipage}

\end{center}

\newpage
\setcounter{page}{1}
\renewcommand{\thefootnote}{\arabic{footnote}}
\setcounter{footnote}{0}

\tableofcontents
\section{Introduction}

Integrability has proven to be a powerful tool for studying the planar ${\mathcal{N}}=4$ SYM theory. 
In particular, it was successfully used to compute all the 
two-point functions of the gauge-invariant single-trace
operators for any value of the 't Hooft parameter $\lambda$, see for instance  \cite{Spectrum,BigReview,pmu}. 
The predictions from integrability have been extensively tested and they correctly reproduce the known results obtained in perturbation theory at weak coupling and the ones obtained by the AdS/CFT conjecture in the strong coupling limit.

The natural next step is computing the three-point functions. Together with the two-point functions these are the building blocks for all the higher point correlators. 
With the help of table \ref{stateofart}, let us briefly recall the state of the art concerning the computation of the three-point functions at weak coupling and explain where our findings fit within this picture. 
\begin{table} 
\begin{centering}
\begin{tabular}{l|l|l|l|l} 
\bf{Sector} & \begin{tabular}{l} \bf{Tree-level and} \\ \bf{Integrability} \end{tabular} &  \begin{tabular}{l}\textbf{One-loop} \\ \bf{prescription} \end{tabular}& \begin{tabular}{l} \bf{One-loop and} \\ \bf{Integrability} \end{tabular}& \begin{tabular}{l} \bf{Higher} \\\bf{loops} \end{tabular} \\ \hline
$\alg{su}(2)$ & \cite{TailoringI},\cite{FodaMetodo} & \cite{Japas},\cite{Alday} & \cite{TailoringIV}, \cite{Kostovmorph} & unknown\\ \hline
$\alg{sl}(2)$ & \cite{Pedrosl2},\cite{Kazakov3ptsl2}& \cite{Alday} \small{\,(some cases)} & \cite{Pedrosl2} \small{\,(some cases)} & unknown\\ \hline
$\alg{su}(1|1)$& {\bf{here}} &{ \bf{here} }& { \bf{here} } & unknown  \\ \hline
$\alg{so}(6)$ & \cite{FodaNovo} \small{\,(some cases)} & \cite{Japas},\cite{Alday} & unknown & unknown \\ \hline 
$\alg{psu}(2,2|4)$ & unknown & unknown  & unknown & unknown \\ \hline
\end{tabular} \nonumber 
\end{centering}
\caption{The current status of the computation of three-point functions.\label{stateofart}}
\end{table}

A single-trace operator of $\mathcal{N}=4$ SYM is thought of as a closed spin chain state. To leading order in the 't Hooft coupling  these spin chain states are very well understood and given by the so-called Bethe ansatz.
The problem at tree-level is purely combinatorial and amounts to \emph{cutting} and \emph{sewing} such spin chains. At the end of the day, this boils down to a computation of some scalar products of Bethe states. Nevertheless, this is a very rich and non-trivial problem. For instance, scalar products between Bethe states in higher rank algebras are not known. It is therefore so far unclear how to perform the computation of the most general $\alg{psu}(2,2|4)$ correlators as indicated by the last row of  table \ref{stateofart}.
\\
\indent This motivates one to start studying the rank one sectors in a systematic way. They consist of the $\alg{su}(2)$, $\alg{su}(1|1)$ and $\alg{sl}(2)$ sectors and they played a very important role in the spectrum problem, see for instance \cite{StaudacherTwoloop}. The first three rows of the table \ref{stateofart} summarize the current knowledge on these sectors. In the $\alg{su}(2)$ case, the final result for the structure constants turns out to be given 
in terms of determinants depending on three sets of numbers called Bethe rapidities while in the $\alg{sl}(2)$ sector, it was found a formula given in terms of a sum over partitions of these Bethe rapidities.  In this paper, we will study the remaining rank one sector.\\
\indent 
Whichever sector we consider, there are, at one-loop, two effects that need to be taken into account. \\
\indent Firstly, there is the two-loop correction to the Bethe state, which is of order $\lambda$ and thus contributes to the one-loop structure constant. This amounts to correct not only the $S$-matrix but also modifying the Bethe ansatz itself by introducing the so-called contact terms. These are required due to the long-range nature of the dilatation operator which couples non-trivially neighboring magnons on the spin chain.
In this regard, some surprises were found recently. The contact terms were found to be ultra-local in $\alg{sl}(2)$ and much simpler than in the $\alg{su}(2)$ case. In this paper we find a remarkably simple form for the contact terms in $\alg{su}(1|1)$ allowing us to fully construct the two-loop Bethe state for an arbitrary number of magnons. \\
\indent Secondly, there is the perturbative correction from the Feynman diagrams. This can be effectively described by an insertion of an operator at the splitting points of the spin chain and this is what we call the \textit{prescription} for the one-loop computation. 
So far, the prescription was only 
fully computed for the $\alg{so}(6)$ sector. For the $\alg{sl}(2)$ sector, partial results were obtained in \cite{Alday} but the complete computation remains to be done. In this paper, we provide the complete one-loop prescription for the $\alg{su}(1|1)$ sector.\\
\indent 

In the end, combining both 
loop contributions for $\alg{su}(1|1)$, we found a strikingly simple formula for the one-loop structure constant $C_{123}$. 
Given three operators $\mathcal{O}_i$ with $N_i$ excitations with momenta 
$\{p^{(i)}_j\}_{j=1}^{N_i}$, and length $L_i$ (the details of the exact setup will be given below), we have
\<
C_{123}=\mathcal{C}\;\frac{\prod\limits_{i=1}^{3}
\prod\limits_{\substack{j< k}}^{N_i}\mathfrak{f}(y^{(i)}_j,y^{(i)}_k)}{\prod\limits_{i=1}^{N_1}\prod\limits_{j=1}^{N_2}\mathfrak{f}(y^{(1)}_i,y^{(2)}_j)}\prod\limits _{k=1}^{N_1} \left[1-(y^{(1)}_{k})^{L_2}\prod\limits_{i=1}^{N_2} \left(-S(y^{(2)}_i,y^{(1)}_k)\right)\right]\,,\label{great}
\>
where $y_j^{(i)}\equiv e^{i p^{(i)}_j}$, $\mathcal{C}$ is a simple normalization factor given in 
(\ref{normalfactor}), $S$ is the $\alg{su}(1|1)\,S$-matrix. The most essential ingredient and main result of this work is the function $\alg{f}$ which is simply given by
\[
\mathfrak{f}(s,t)=(s-t)\left[1-\frac{g^2}{2}\left(\frac{s}{t}+\frac{t}{s}-\frac{1}{s}-s-\frac{1}{t}- t +2\right)+\mathcal{O}(g^4)\right]\,,
\]
with $g^2= \frac{\lambda}{16 \pi^2}$.  

The paper is organized as follows. In section 2, we explain the three-point function setup that will be used in the remaining of the paper and compute the leading contribution to the structure constants in terms of a simple expression which is function of the momenta of the excitations. Section 3 is devoted to the calculation of the one-loop corrected structure constants.
The section begins with the construction of the two-loop eigenstates by computing the contact terms,  then we evaluate the relevant Feynman diagrams needed for determining the prescription for computing the one-loop corrections. In the end, we put the different contributions together and we arrive at the formula (\ref{great}). Finally, the section 4 contains our 
conclusions and perspectives. Several Appendices have additional details omitted during the presentation.

\section{Three-point functions at leading order}\label{setup}

In this section, we perform the computation of the structure constants at leading order.  The setup that will be used for the calculation involves composite operators made out of  both fermionic and scalar fields. 
Each of these operators is thought of as a state of a closed spin chain with the fermionic fields being excitations over a ferromagnetic vacuum. The advantage of this approach is that the connection with the integrability tools of quantum spin chains   becomes manifest (see for instance \cite{BigReview}) and facilitates the combinatorial problem.

The smallest (closed) sector of $\mathcal{N}=4$ SYM containing both fermionic and bosonic fields is the $\alg{su}(1|1)$. The field content of this sector consists of one complex scalar that we will denote as $Z=\Phi^{34}$ and a complex chiral fermion that shares with the scalar one $R$-charge index, for instance $\Psi = \psi^{4}_{\alpha=1}$. 
The setup for the calculation of the planar three-point functions that we will be considering involves an operator $\mathcal{O}_1$ given by a linear combination of single traces made out of products of these fields. More precisely,
\[
\mathcal{O}_1 =  \sum_{1 \le n_1 < n_2 < \ldots < n_{N_1} \le L_1} \psi^{(1)}(n_1, n_2, \ldots, n_{N_1})   \Tr\left( Z \dots \underset{n_1} \Psi \dots \underset{n_2}\Psi \dots Z\right)  \label{defO1}\, , 
\]
where $L_1$ is the length of the operator, $N_1$ is the number of its fermionic fields and 
$n$'s are the positions of the excitations along the chain of $Z$'s. We designate the coefficients $\psi^{(1)}$ in this linear combination 
by wave-function. It is natural to consider the second operator $\mathcal{O}_2$ made out of the complex conjugate fields, namely
\<
\mathcal{O}_2 =  \sum_{1 \le n_1 < n_2 < \ldots < n_{N_2} \le L_2} \psi^{(2)}(n_1, n_2, \ldots, n_{N_2})   \Tr\left( \bar{Z} \dots \underset{n_1}{\bar\Psi} \dots \underset{n_2}{\bar\Psi} \dots \bar{Z}\right)  \,. \label{defO_2}
\>
In our conventions, the complex conjugate fields are given by
\<
\bar{Z} = (Z)^{*}=\Phi_{34} = \Phi^{12} \, , \\ 
\bar{\Psi} = (\Psi)^{\dag}=\bar{\psi}_{4,\, \dot{\alpha}=\dot{1}} \, . \hspace{7.5mm} \nonumber
\>
From now on, we will omit the Lorentz spinorial indices at several places keeping in mind that they are always kept fixed.
\begin{figure}[t]
\centering
\includegraphics[width=0.39\linewidth]{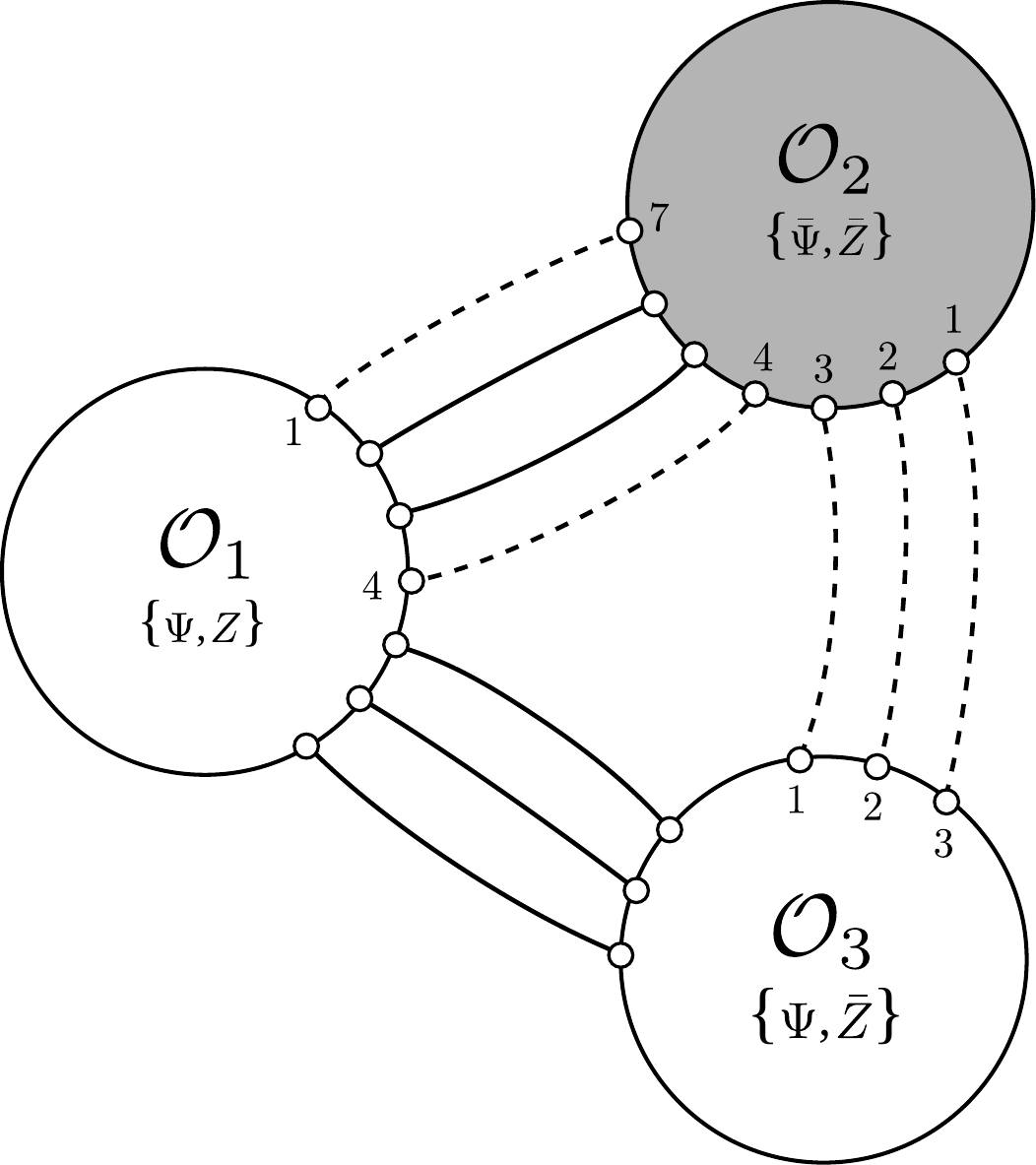}
\caption{The leading order contribution to the three-point functions. 
The solid lines represent a bosonic propagator and the dashed lines represent a fermionic propagator. We also indicate our conventions for labeling the positions of the excitations. Notice that in our setup the first $N_3$ excitations of the operator $\mathcal{O}_2$ have always their position fixed. \label{Thesetupandorder}}
\end{figure}

As a consequence of the $R$-charge conservation, it is clear that we cannot take the third operator to be also in the same $\alg{su}(1|1)$ sector to which $\mathcal{O}_1$ and $\mathcal{O}_2$ belong, if we want to have a non-vanishing result and avoid extremal correlation functions\footnote{The extremal case presents additional subtleties related to the mixing with double-trace operators, see \cite{TailoringI}. We will not investigate such issues in this paper and therefore only non-extremal three-point functions will be considered.}.  Instead, we consider a ``rotated'' operator constructed by applying $\alg{su}(4)$ generators several times to a $\alg{su}(1|1)$ operator of the type $\mathcal{O}_1$. 
The idea is to get a composite operator having a term with only   $\Psi$ and $\bar{Z}$ fields in order to allow non-vanishing Wick contractions between all pairs of operators (see figure \ref{Thesetupandorder} for an example of a non-extremal three-point function). 
More precisely, let us suppose that we start with a state made out of $\Psi$ and $Z$ fields. In order to convert a single $Z$ into a $\bar{Z}$ we must apply a pair of $\alg{su}(4)$ generators that rotate its two $R$-charge indices. In sum, we can generate a term with $\Psi$'s and $\bar{Z}$'s by considering the following operation
\[
\mathcal{O}_3 = \frac{1}{ (L_3-N_3)!^2}(\alg{R}^{2}_{\;\;4 } \alg{R}^{1}_{\;\;3 })^{L_3-N_3}    \sum_{1 \le n_1 < \ldots < n_{N_3} \le L_3} \psi^{(3)}(n_1, \ldots, n_{N_3})   \Tr\left( Z \dots \Psi \dots \Psi \dots Z\right)\,, \label{def03}
\]
where $\alg{R}^{a}_{\;\;b}$ are $\alg{su}(4)$ generators and they act on the fields inside the trace. Now, the $\alg{su}(4)$ generators may also act on the field $\Psi$ which carries one $R$-charge index. Therefore, this operation will generate several terms coming from the different ways of acting with the generators,
\< 
\mathcal{O}_3 \,=\,  \sum_{1 \le n_1 < \ldots < n_{N_3} \le L_3} \psi^{(3)}(n_1, \ldots, n_{N_3})  \Bigl[&&\Tr\left( \bar{Z} \ldots \Psi \dots \Psi \ldots  \bar{Z}\right)+ \nonumber \\
&+& \,\,\Tr \left( \bar{Z} \dots  \psi^2 \dots \Psi \dots \Phi^{14}\, \right) +\ldots\, \Bigr] \label{eq:rotation}  \, , 
\>
where in the first line we have the term where all the $\alg{su}(4)$ generators act on the scalar fields $Z$. In the second line, we represent the terms where some of the generators also act on the fermionic fields $\Psi$.
As an example of how the formula given above is evaluated consider,
\[ \label{rotation}
(\alg{R}^{2}_{\;\;4 } \alg{R}^{1}_{\;\;3 }) \cdot \text{Tr} \, (\Psi Z)= (\alg{R}^{2}_{\;\;4 } \alg{R}^{1}_{\;\;3 })\cdot \text{Tr} \, (\psi^4 \, \Phi^{34} ) = \text{Tr} \, ( \psi^2 \, \Phi^{14}) + \text{Tr} \, (\psi^4 \, \Phi^{12})  \, . \nonumber
\]

At tree-level, the terms in the second line of (\ref{eq:rotation}) do not give any contribution due to the $R$-charge conservation. In other words,
one always has a zero Wick-contraction. Therefore, at leading order, only the first line contributes and we get a tree-level diagram of the type represented in figure \ref{Thesetupandorder}. At one-loop, the terms in the second line will also need to be taken into account. We emphasize that the operators $\mathcal{O}_1$ in (\ref{defO1}) and $\mathcal{O}_3$ in (\ref{def03}) are spinorial operators with $N_1$ and $N_3$ indices $\alpha=1$ respectively. This follows from the definition of the field $\Psi$ given previously. The operator $\mathcal{O}_2$ in  (\ref{defO_2}) has $N_2$  Lorentz indices $\dot\alpha=\dot{1}$ associated to each of the fermions $\bar{\Psi}$.   \\

\indent In a conformal field theory,  the two-point functions are completely fixed by the symmetries up to a normalization constant. For two operators having spinorial indices as shown below, we have
\begin{equation}
\langle \mathcal{O}_{i \, ; \, 1_1 ... 1_{N_i}}(x_1) \, \bar{\mathcal{O}}_{i \, ; \, \dot{1}_1 ... \dot{1}_{N_i}}(x_2) \rangle =  \mathcal{N}_i \, \frac{ (J_{12, 1 \dot{1} })^{N_i}}{|x_{1 2}|^{2 \Delta_i}} \label{2ptstruct}\, ,
\end{equation}
where $\mathcal{N}_i$ is a constant associated to the normalization of the operator, $\Delta_i$  is its conformal dimension and the tensorial structure is\footnote{See Appendix \ref{appnotations} for our conventions.}
\begin{equation}
J_{ij, 1 \dot{1}} =  \frac{ x_{ij}^{\mu} \left(\sigma^{E}_{\mu}\right)_{1 \, \dot{1}} }
{\left(2 \pi \right)^2 |x_{ij}|}  \, , \quad \quad \text{with} \quad \quad x_{ij}^{\mu} = x^{\mu}_i - x^{\mu}_j \, . \label{Jtwo-point}
\end{equation}

In the case of three-point functions of generic operators having spinorial indices, one has many inequivalent 
tensor structures consistent with the conformal symmetry, and the result of the correlation function is a linear combination of these structures. The constraints following from conformal symmetry on the higher point functions were studied for instance in \cite{Sotkov1,Sotkov2,Spinning}.  However,  for the setup considered in this work there is 
only one possible tensor and the three-point functions is of the form
 
\begin{eqnarray}
\langle \mathcal{O}_{1  ;  1_1 \dots 1_{N_1}}(x_1) \,  \mathcal{O}_{2  ;  \dot{1}_1 \dots \dot{1}_{N_2}}(x_2) \, \mathcal{O}_{3  ;  {1}_1 \dots {1}_{N_3}}(x_3) \rangle = \hspace{50mm} \\ 
\frac{ (J_{12,1 \dot{1}})^{N_1}  (J_{23,1 \dot{1}})^{N_3} \,  \sqrt{ \mathcal{N}_1 \mathcal{N}_2 \mathcal{N}_3} \, C_{123} (g^2)  }{|x_{12}|^{\Delta_1 +\Delta_2 -\Delta_3}|x_{13}|^{\Delta_1 +\Delta_3 -\Delta_2}|x_{23}|^{\Delta_2 +\Delta_3-\Delta_1}} \,, \nonumber
\end{eqnarray}
where we are considering $N_2 = N_1 + N_3$ 
and  $g^2=\frac{\lambda}{16\pi^2}$ with $\lambda$  the  't Hooft parameter. 

\indent The structure constant $C_{123}(g^2)$ has a perturbative expansion when $g^2$ is small, and its leading order will be designated by $C_{123}^{(0)}$. Using the figure \ref{Thesetupandorder}, we observe that the only non-trivial Wick contractions occur between operators $\mathcal{O}_1$ and $\mathcal{O}_2$. The structure constant $C_{123}^{(0)}$ is then given by the product of the three wave-functions with a sum over the positions of the excitations between these two operators,
\begin{eqnarray}
\left|  C^{(0)}_{123}  \right| =  \alpha 
\left|  \psi^{(3)}_{1, \ldots, N_3} \sum_{N_3< n_1 < \ldots < n_{N_1} \le L_2 } \psi^{(1)}_{L_2+1-n_{N_1}, \ldots, L_2 + 1 - n_1}   \psi^{(2)}_{1, \ldots, N_3,  n_1, \ldots, n_{N_1}}\right| \, \label{leadingorder} . 
\end{eqnarray}
$\alpha$ is a normalization factor that comes from the fact that we are normalizing the operators such that their two-point functions has the canonical form ($\ref{2ptstruct}$) with $\mathcal{N}_i=1$. It is given by
\begin{equation}
\alpha = \sqrt{ \frac{ L_1 L_2 L_3}{{\mathcal{N}}^{(1)} {\mathcal{N}}^{(2)} {\mathcal{N}}^{(3)}}} \, , \;\;\; \text{with}\;\;\; {\mathcal{N}}^{(j)} = \sum_{1 \le n_1 < \ldots < n_{N_j} \le L_{j}} ( \psi^{(j)}_{n_1, \ldots, n_{N_j}} )^{*} (\psi^{(j)}_{n_1, \ldots, n_{N_j}}) \, . \label{norm}
\end{equation}
The main goal of this section is to find a closed formula for $C_{123}^{(0)}$.

\subsection{The one-loop Bethe eigenstates and structure constants}

To compute $C_{123}^{(0)}$ we must consider states with definite one-loop anomalous dimension \cite{TailoringI}. The one-loop $\alg{su}(1|1)$ integrable Hamiltonian and  $S$-matrix can be found in \cite{BeisertTwoloop,StaudacherTwoloop}. The Hamiltonian is simply the fermionic version of the Heisenberg Hamiltonian and 
it is written in terms of the Pauli matrices as
\begin{equation}
H_1=2 g^2\sum_{n=1}^{L}\left((1-\sigma_n^{3})-\frac{1}{2}(\sigma_{n}^{1}\sigma_{n+1}^{1}+
\sigma_{n}^{2}\sigma_{n+1}^{2})\right) \label{eq:H1} \, ,
\end{equation} 
where $L$ is the length of the spin chain. At leading order the two-excitation $S$-matrix 
is independent of their momenta and simply given by
\begin{equation}
S
(p_1,p_2)=-1 \, . 
\end{equation}

In order to find the eigenstates of the Hamiltonian given above, we 
use the usual coordinate Bethe ansatz.
A $N$-magnon state of a spin-chain of length $L$ is of the form 
\begin{equation}
| \psi_N \rangle = \sum_{1 \le n_1 < n_2 < \ldots < n_N \le L} \psi_N(n_1, n_2, \ldots, n_N) |n_1, \ldots, n_N \rangle \, , \label{eq:coordBethe}
\end{equation}
where the $n_i$'s  in $|n_1, \ldots \, n_N \rangle$ indicate the position of the fermionic excitations $\Psi$ on the
chain  (for details about the coordinate Bethe ansatz see  \cite{BigReview,TailoringI}). Notice that the ket  $|n_1, \ldots \, n_N \rangle$ represents the trace in (\ref{defO1}).
The wave-function $\psi_N(n_1, \ldots, n_N)$ is a combination of plane waves with as many terms as the number of possible permutations of the momenta with the relative coefficients 
being the $S$-matrices. Since the leading order $\alg{su}(1|1)$ $S$-matrix is just $-1$, the several terms in the wave-function will appear with alternating signs which we write as
\[
\psi_N(n_1, n_2, \ldots, n_N)= \sum_{P}\, \text{sign}\,P \;{{\rm exp}} (i p_{\sigma_P(1)} n_1 + i p_{\sigma_P(2)} n_2 + \ldots + i p_{\sigma_P (N)} n_N) \label{eq:wave-functioncoordinate}
\]
where $P$ indicates sum over all possible permutations $\sigma_P$ of the elements $\{1, \ldots, N\}$, and  $\text{sign}\,P$ is the sign of the permutation. 
Moreover, we should impose the periodicity condition by requiring the momenta $p_i$ to satisfy the Bethe equations  
\begin{equation}
e^{i p_i L}=1\,. \label{eq:Betheequations}
\end{equation}
The cyclic property of the trace is implemented by imposing the zero momentum condition 
of the state,
\begin{equation}\label{totalm}
\sum_{i=1}^{N} \, p_i = 2\pi \times \text{integer} \, .
\end{equation}

Having determined the eigenstates of the one-loop $\alg{su}(1|1)$ Hamiltonian, we can proceed to compute the leading order structure constant $C^{(0)}_{123}$ given in (\ref{leadingorder}) by following some simple steps. First, we notice that since the positions of the excitations of the third operator are fixed, we can use (\ref{eq:wave-functioncoordinate}) to write $\psi^{(3)}$  explicitly. It is simple to see that we obtain a Vandermonde determinant which can be also presented as a simple product, 
\begin{eqnarray} \label{sum}
\left|  C^{(0)}_{123}  \right| =  \alpha \left|\,  \prod\limits_{j<k}^{N_3}\left[e^{i p^{(3)}_j }-e^{i p^{(3)}_k} \right]
 \sum_{N_3< n_1 < \ldots < n_{N_1} \le L_2 } (\psi^{(1)}_{n_{1}, \ldots, n_{N_1}})^{*} \, \psi^{(2)}_{1, \ldots, N_3,  n_1, \ldots, n_{N_1}}   \, \right| \, . \hspace{4mm}
\end{eqnarray}
Moreover we have replaced $\psi^{(1)}_{L_2+1-n_{N_1}, \ldots, L_2 + 1 - n_1} $ by $ (\psi^{(1)}_{n_{1}, \ldots, n_{N_1}})^{*}$ since they differ by at most a sign.

Notice that the first $N_3$ excitations of the wave-function $\psi^{(2)}$ have their positions fixed or \textit{frozen}. In order to make the computation of this sum simpler, we consider an auxiliary problem where we add $N_3$ extra excitations to the wave-function $\psi^{(1)}$ and liberate the fixed $N_3$ roots of $\psi^{(2)}$ with their positions being summed over too,
\<
\mathcal{S}_{aux}\,\equiv\, \sum_{1\leq n_1<  \ldots < n_{N_3+N_1} \le L_2 } (\psi^{(1)}_{n_1, \ldots, n_{N_3 + N_1}})^{*} \, \psi^{(2)}_{n_1, \ldots, n_{N_3+N_1}} . \label{aux}
\>
The advantage of considering this auxiliary problem is that the sum (\ref{aux}) 
can be easily computed due to the form of the wave-functions.
Moreover, we can relate it with the original sum appearing in (\ref{sum}) as we now explain. Indeed, let us consider that $N_3$ momenta, say $\{ p^{(1)}_{1},\ldots, p^{(1)}_{N_3} \}$, are complex. We can then dynamically localize the wave-function around the original $N_3$ positions by taking the limit of these momenta going to minus infinity. More precisely, we send
$\{e^{- i p^{(1)}_{1}},\dots,e^{- i p^{(1)}_{N_3}}\}$ 
to zero in such a way that
\[
e^{- i p^{(1)}_{1}} \ll \dots \ll e^{- i p^{(1)}_{N_3}} \label{order} \,.
\]
Thus, given the explicit form of the wave-function (\ref{eq:wave-functioncoordinate}), we observe that in this limit the sum over the positions of the extra roots in (\ref{aux}) is dominated by the term for which $n_1=1,\dots,n_{N_3}=N_3$. This procedure of sending roots to a particular limit in order to freeze their positions is the coordinate Bethe ansatz counterpart of the \textit{freezing trick} used in \cite{FodaMetodo} at the level of the six-vertex model. Neglecting all the subleading terms, we get that in this limit, (\ref{aux}) is reduced to 
\[
\mathcal{S}_{aux} \rightarrow\left(\prod_{k=1}^{N_3}e^{-i p^{(1)}_{k} \, k}\right)  \sum_{N_3< n_1 < \ldots < n_{N_1} \le L_2 } (\psi^{(1)}_{n_{1}, \ldots, n_{N_1}})^{*} \, \psi^{(2)}_{1, \ldots, N_3,  n_1, \ldots, n_{N_1}} , \label{aux2}
\]
where we recognize precisely the original sum of  (\ref{sum}).\\
\indent Returning to our auxiliary problem, we use again that the wave-function is completely antisymmetric in its arguments to extend the limits of the sum (\ref{aux}). In compensation, we merely have to introduce a trivial overall combinatorial factor. 
Using the explicit form of the wave-function we write the sum (\ref{aux}) as
\[
\mathcal{S}_{aux}\,=\,\frac{1}{N_2!}\sum_{\{n_i\}}\sum_{P, Q}\, \text{sign}\, P\,  \text{sign}\,Q \, \prod_{a=1}^{N_1+N_3} e^{ \left(ip^{(2)}_{P(a)}-ip^{(1)}_{Q(a)}\right) n_a}\,. \label{aux4}
\]
We emphasize again that we now sum without restrictions, $1\leq n_i \leq L_2 $, for all  $n_i$.
These sums over $n_i$ can be explicitly computed as they are geometric series. Using the Bethe equations and the total momentum condition for the operator $\mathcal{O}_2$, we can then simplify (\ref{aux4}) to
\[
\mathcal{S}_{aux}\,=\,\left[\prod_{a=1}^{N_1+N_3}\left(1-e^{-i p^{(1)}_{a}L_2}\right)\right]\frac{1}{N_2!} \sum_{P, Q}\, \text{sign}\, P\,  \text{sign}\, Q \, \prod_{a=1}^{N_1+N_3}  \frac{1}{e^{i p^{(1)}_{Q(a)}}-e^{i p^{(2)}_{P(a)}}}\,.
\]
The remaining sum in the previous expression is manifestly the definition of 
a Cauchy determinant and, therefore, it can be 
written explicitly as a simple product as follows
\[
\mathcal{S}_{aux}\,=\,\left[\prod_{a=1}^{N_1+N_3}\left(1-e^{-i p^{(1)}_{a}L_2}\right) \right] \, \frac{\prod\limits_{j<k}(e^{i p^{(1)}_{j}}-e^{i p^{(1)}_{k}})(e^{i p^{(2)}_{k}}-e^{i p^{(2)}_{j}})}{\prod\limits_{j,k}(e^{i p^{(1)}_{j}}-e^{i p^{(2)}_{k}})}\, \label{sumfinal}.
\]
Notice that this expression contains as a limit the norm of an operator.\footnote{If we set $N_3=0$ and consider $p^{(1)}_j\rightarrow p^{(2)}_j$ we get the expression for $\mathcal{N}^{(2)}$ after using the Bethe equations (\ref{eq:Betheequations}).} It is given by 
\[
\mathcal{N}^{(j)} = L_j^{N_j}\,.
\]
\indent Finally, we take the limit of (\ref{sumfinal}) when 
$\{e^{-i p^{(1)}_{1}},\dots,e^{-i p^{(1)}_{N_3}}\}$ 
vanish as in (\ref{order}).
Plugging the resulting limit and taking into account the overall product multiplying the sum in (\ref{aux2}), we obtain our final result
\[
\left| C^{(0)}_{123}  \right| =\left[\prod\limits_{i=1}^{3}L_i^{\frac{1-N_i}{2}}\right]  \left| \left[\prod_{j=1}^{N_1}\left(1-e^{i p^{(1)}_{j}L_2}\right)\right] \, \frac{\prod\limits_{a=1}^{3}\prod\limits_{j<k}^{N_a}(e^{i p^{(a)}_{j}}-e^{i p^{(a)}_{k}})}{\prod\limits_{j=1}^{N_1}
\prod\limits_{k=1}^{N_2}(e^{i p^{(1)}_{j}}-e^{i p^{(2)}_{k}})}\, \right| \label{finaltree}\,. 
\]

It is now straightforward to confirm that our formula (\ref{great}) given in the introduction, reduces to this one when $g$ is set to zero.\\
\indent This result fills the first column for the $\alg{su}(1|1)$ row of the table \ref{stateofart} in the introduction. Let us remark that this expression is considerably simpler than the ones found for the $\alg{su}(2)$ and $\alg{sl}(2)$ sectors. This is perhaps not surprising given that at leading order we are dealing with a theory of free fermions so that the form of the $\alg{su}(1|1)$ wave-function becomes quite simple. However, we will see that the one-loop result persists to be simpler than in the other sectors.

\section{One-loop three-point functions}
In this section, we compute the structure constants at first order in the 't Hooft coupling $\lambda$ for our setup. There are two main ingredients in this computation. Firstly, one has to consider Bethe eigenstates that diagonalize the two-loop dilatation operator as these states are of order $\lambda$. Secondly, one has to compute the relevant Feynman diagrams at this order in perturbation theory. This second contribution can be compactly taken into account through the insertion of an operator at specific points of the spin chains as will be reviewed.

\subsection{Two-loop coordinate Bethe eigenstates and Norms}\label{two loop}
The two-loop Bethe eigenstates are determined by diagonalizing the long-range Hamiltonian $H$ \cite{StaudacherTwoloop}
\begin{equation}
H = H_1 + H_2 \, , \label{eq:twoloophamiltonian}
\end{equation}
where $H_1$ is given in (\ref{eq:H1}) and 
\<
H_2= 4 g^2 \sum_{n=1}^{L} \Bigl( 2(\sigma_n^{3}-1)-\frac{1}{4}(\sigma_{n}^{3} \sigma_{n+1}^{3}-1)+(\sigma_{n}^{1}\sigma_{n+1}^{1}+\sigma_{n}^{2}\sigma_{n+1}^{2})\left( \frac{9}{8}-\frac{1}{16}\sigma_{n+2}^3\right) \hspace{5mm} \\
   -\frac{1}{16}\sigma_{n}^{3}(\sigma_{n+1}^{1}\sigma_{n+2}^{1}+\sigma_{n+1}^{2}\sigma_{n+2}^{2})-\frac{1}{8}\sigma_{n}^{1}(1+\sigma_{n+1}^{3})\sigma_{n+2}^{1}-\frac{1}{8}\sigma_{n}^{2}(1+\sigma_{n+1}^{3})\sigma_{n+2}^{2}\Bigr)\,,  \nonumber
\>
where $\sigma^i$ are the Pauli matrices. In order to diagonalize it, we start with the usual coordinate Bethe ansatz which works when the excitations are at a distance bigger than the range of the interaction, i.e. when $|n_i-n_j|>2$. In this region all we need is the two-loop 
$S$-matrix which reads
\[
S(p_1,p_2)=-1-8 i g^2 \sin \left( \frac{p_1}{2}\right)\sin \left( \frac{p_1-p_2}{2}\right)\sin \left( \frac{p_2}{2}\right)\,. \label{2loopSm}
\]
Given the long-range nature of the Hamiltonian (\ref{eq:twoloophamiltonian}), we expect the form of the wave-function to be modified with respect to the usual Bethe ansatz (\ref{eq:wave-functioncoordinate}). In fact, when magnons are placed at neighboring positions on the spin chain they interact in a non-trivial way. Therefore, the wave-function must be refined by the inclusion of the so-called \textit{contact terms}. For instance, in the case of three magnons we write it as
\<
\psi(n_1,n_2,n_3)=\phi_{123}+\phi_{213}S_{21}+
\phi_{132}S_{32}+
\phi_{312}S_{31}S_{32}
+\phi_{231}S_{31}S_{21}+\phi_{321}S_{32}S_{31}S_{21}  \, ,  \nonumber
\>
where we have used the notation $S_{ab}=S(p_a,p_b)$ and 
\<
\phi_{abc}=e^ {i p_a n_1 +i p_b n_2 +i p_c n_3}\Bigl(&1&+g^2\, \mathbb{C}(p_a,p_b)\,\delta_{n_2,n_1+1}\delta_{n_3>n_2+1}+g^2 \mathbb{C}(p_b,p_c)\,\delta_{n_2>n_1+1}\delta_{n_3,n_2+1}\nonumber\\
&+&g^2 \,\mathbb{C}(p_a,p_b,p_c)\,\delta_{n_2,n_1+1}\delta_{n_3,n_2+1}\Bigr) \,.
\>
The functions $\mathbb{C}$ are the contact terms which are fixed by solving the energy eigenvalue problem.
 In the case of $N$-magnons, the wave-function has a similar structure. It consists of $N!$ terms coming from the permutations of $\{p_1,\dots,p_N\}$ and $N-1$ types of contact terms namely $\mathbb{C}(p_i,p_j),\dots, \mathbb{C}(p_1,...,p_N)$. \\
\indent Unexpectedly, we have found that up to seven magnons the contact terms are simply given by\footnote{We thank Tianheng Wang for collaboration on this point.} 
\[
\mathbb{C}(p_1,\dots,p_N)=\frac{N-1}{2} \label{cterms}\,.
\]
Even though we have not proved the validity of this formula for an arbitrarily high number of magnons, the pattern emerging up to seven magnons is quite suggestive.
Given the form of the contact terms in the $\alg{su}(2)$ and $\alg{sl}(2)$ sectors, the simplicity of the $\alg{su}(1|1)$ result is quite surprising. In particular, notice that they are independent of the momenta of the colliding magnons. This might be pointing towards the existence of a new algebraic description of these states yet to be unveiled.

\indent As already explained, in order to correctly compute the three-point functions we need to know the norm of the Bethe eigenstates as we are normalizing the result by the two-point functions. Remarkably, we have checked numerically up to six-magnons that the two-loop (coordinate) norm is given by 
\[
\mathcal{N}
=   \det_{j,k\le N} \frac{\partial }{\partial p_j} \left[ L p_{k}
+\frac{1}{i}\sum_{m\neq k}^{N} \log S(p_m,p_k) \right] \,\label{normsu11} .
\]
Interestingly, this formula is precisely the well-known Gaudin norm for the one-loop $\alg{su}(2)$ Bethe states. Still within the $\alg{su}(2)$ sector, it was recently shown in \cite{TailoringIV} that this expression remains valid at higher loops leading to an all-loop conjecture for the norm. Moreover, the two-loop norm for $\alg{sl}(2)$ Bethe states was found to be precisely of the type (\ref{normsu11}) as described in \cite{Pedrosl2}. In all these cases, the contact terms recombine exactly to preserve the determinant form. This is very suggestive of an underlying hidden structure that is worth investigating.

\subsection{One-loop perturbative calculation}
\indent Loop computations will 
give rise to divergences which require the introduction of a regularization scheme. A very convenient one and the one that will be used in this work is the \textit{point splitting} regularization. At one-loop, only neighboring fields inside any of the single-trace operators interact and the divergences arise because the two fields are at the same spacetime point. The idea behind the point splitting regularization is to separate these two fields by a distance $\epsilon$ which will act as a regulator
\footnote{In order to preserve the gauge invariance, one can introduce a Wilson line between the two shifted fields. This will in principle introduce extra diagrams at one-loop, coming from the gluon emission from the Wilson line. However, we will show in the Appendix \ref{Wline} that this additional contribution actually vanishes at this order in perturbation theory.}. 

Consider a $\alg{su}(1|1)$ bare operator which is an eigenstate of the one-loop dilatation operator. Its non-vanishing two-point function is of the form  
\begin{equation}
\langle \mathcal{O}_{i \, ; \, 1_1 ... 1_{N_i}}(x_1) \, \bar{\mathcal{O}}_{i \, ; \, \dot{1}_1 ... \dot{1}_{N_i}}(x_2) \rangle =  \mathcal{N}_i \, \frac{(J_{12, 1 \dot{1} })^{N_i}}{|x_{1 2}|^{2 \Delta_{0,i}}}\left(1+2
g^2 
 \, a_i  -\gamma_i \log \left(\frac{x_{12}^2}{\epsilon^2} \right) \right) \, ,
\end{equation}
where the tensor on the right-hand side was defined in (\ref{Jtwo-point}). In the expression above, $\Delta_{0,i}$ and $\gamma_i$ are the free scaling dimension and the one-loop anomalous dimension of the operator $\mathcal{O}_i$ respectively, $\mathcal{N}_i$ is a normalization constant and $a_i$ is a scheme dependent constant. In addition, the three-point function of three $\alg{su}(1|1)$ bare operators that diagonalize the one-loop dilatation operator is, in our setup, fixed by conformal symmetry and takes the form (see \cite{Japas} for details)  
\begin{eqnarray}
\langle \mathcal{O}_{1 \, ; \, {1}_1 ... {1}_{N_1}}(x_1) \, \mathcal{O}_{2 \, ; \, \dot{1}_1 ... \dot{1}_{N_2}}(x_2) \, \mathcal{O}_{3 \, ; \, {1}_1 ... {1}_{N_3}}(x_3) \rangle = \label{eq:3ptbare}
\end{eqnarray}
\begin{eqnarray}
\frac{  (J_{12,1 \dot{1}})^{N_1}  (J_{23,1 \dot{1}})^{N_3}\,\, \sqrt{\mathcal{N}_1 \mathcal{N}_2 \mathcal{N}_3} \, }{|x_{12}|^{\Delta_{0,1} +\Delta_{0,2} -\Delta_{0,3}}|x_{13}|^{\Delta_{0,1} +\Delta_{0,3} -\Delta_{0,2}}|x_{23}|^{\Delta_{0,2} +\Delta_{0,3}-\Delta_{0,1}}}  
 \,  \, C_{123}^{(0)} \, \times \nonumber 
\end{eqnarray}
\begin{eqnarray} 
\left(1+
g^2 
\,  (C_{123}^{(1)}+a_1+a_2+a_3)  -\frac{\gamma_1}{2}\log \left( \frac{x_{12}^2 x_{13}^{2}}{x_{23}^{2}\epsilon^2}\right) -\frac{\gamma_2}{2}\log \left(\frac{x_{12}^2 x_{23}^{2}}{x_{13}^{2}\epsilon^2}\right)-\frac{\gamma_3}{2}\log \left( \frac{x_{23}^2 x_{13}^{2}}{x_{12}^{2}\epsilon^2}\right) 
\right) \hspace{60mm}  \nonumber  
\end{eqnarray}
where we have factored out the tree-level constant
$C_{123}^{(0)}$. 

To extract the regularization scheme independent structure constant $C^{(1)}_{123}$ from the expression above, we have to divide the 
three-point function by the square root of the two-point functions of all the operators to get rid of the constants $a_i$'s. 
After performing this division, one can then read the meaningful structure constant. 

\begin{figure}[t]
\centering
\def\svgwidth{13.0cm}
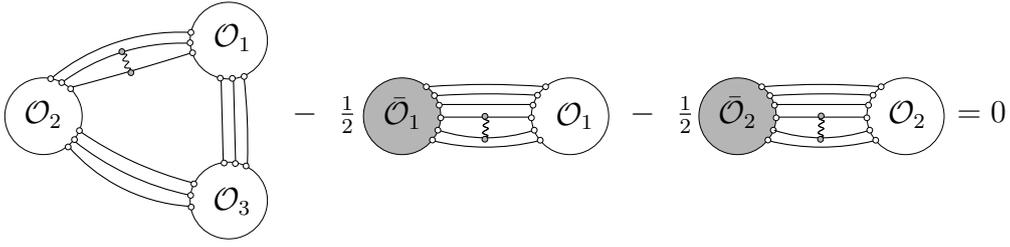
\caption{The wavy-line in the figure is just a representation of a one-loop diagram (for example, a gluon exchange). 
When the contribution of the square root of the two-point functions is  subtracted (this is the reason for the factor $\frac{1}{2}$), all the diagrams involving just two operators are canceled. \label{cancellation}}
\end{figure}

From the Feynman diagrams computation point of view, it is actually simpler to calculate $C_{123}^{(1)}$  instead of the combination $(C_{123}^{(1)}+a_1+a_2+a_3)$. In fact, because we have to divide by the square root of the two-point functions, all one-loop diagrams in the three-point function involving only two operators are canceled. The figure \ref{cancellation} has an example of a such cancellation.

\begin{figure}[t]
\centering
\def\svgwidth{8.0cm}
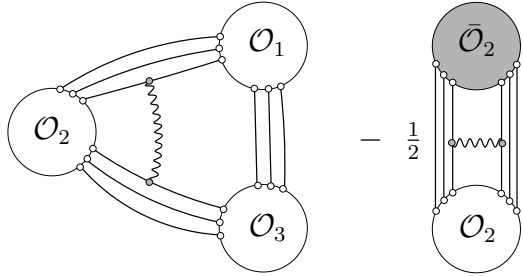
\caption{ A genuine three-point diagram to which we subtract half of the same diagram but seen as a two-point process is shown. The constant coming from this combination of diagrams is regularization scheme and normalization independent. \label{3ptdiagram}}
\end{figure}

\begin{figure}[t]
\begin{center}
\includegraphics[width=0.90\linewidth]{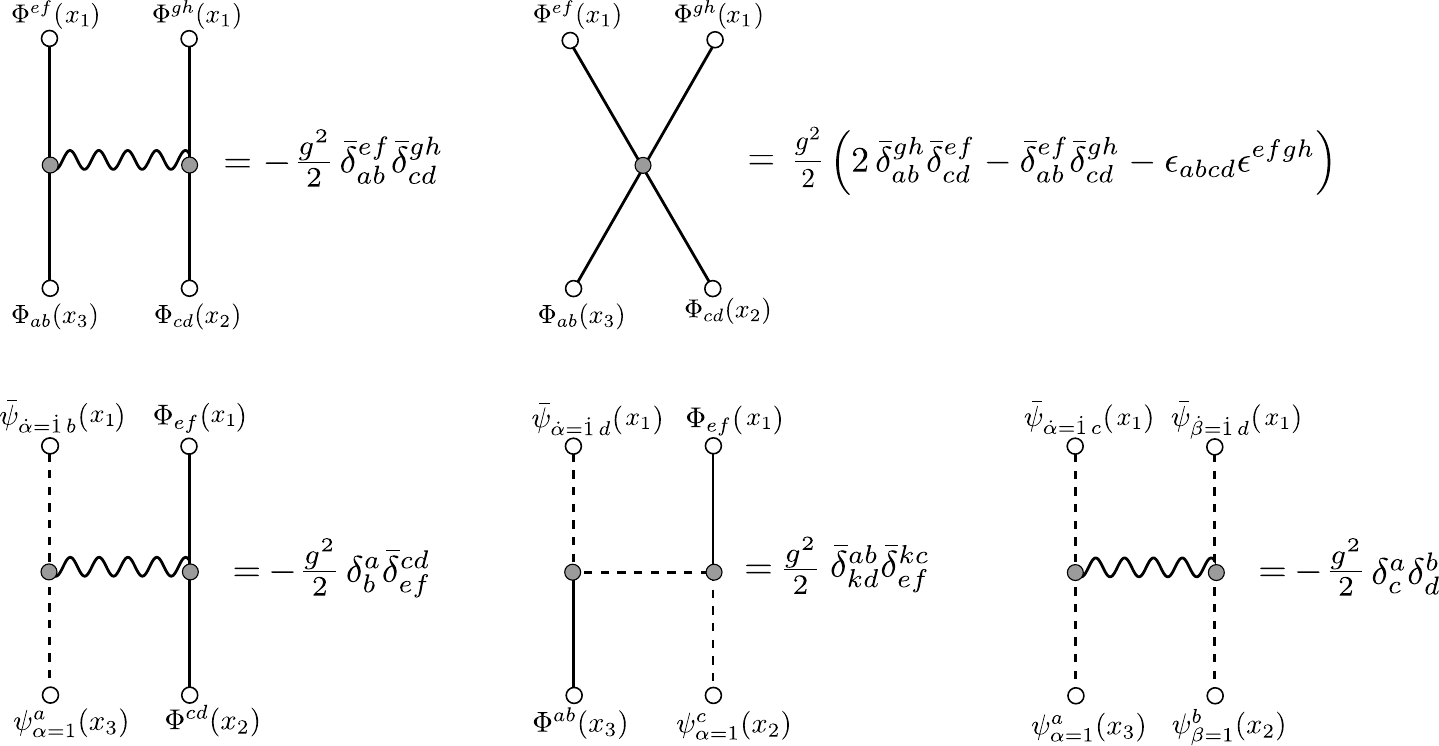}
\caption{These are the relevant one-loop diagrams for the three-point functions. All other graphs give a zero contribution. The solid, wiggly and dashed lines represent the scalars, gluons and fermions, respectively. 
The constants are obtained by combining the three-point and two-point graphs as illustrated in figure \ref{3ptdiagram}. We have used the point splitting regularization and the Feynman gauge. For three-point diagrams we take the limit where a pair of dots (either top or bottom) are brought to the same spacetime points. For the two-point function, both pairs of dots (top and bottom) are brought to the same spacetime points. We are using the definition $\bar\delta_{cd}^{ab}\equiv\delta_{c}^{a}\delta_{d}^{b}-\delta_{c}^{b}\delta_{d}^{a}.$}\label{results}
\end{center}
\end{figure}

The conclusion is that one is left with the computation of only genuine three-point diagrams, i.e., the diagrams involving fields from the three operators\footnote{This fact was dubbed the \textit{slicing argument} in \cite{Alday}}. The allowed positions of the spin chains where it is possible
to have those genuine diagrams  
are commonly called the \textit{splitting points}. 
We are then seeking the constants coming from the genuine three-point diagrams subtracted by the constants coming from the same diagrams but now seen as two-point processes. This is exemplified in the figure \ref{3ptdiagram}.
\\
\indent The details of the Feynman diagram computation are given in the Appendix \ref{appdetails} and here we just provide the results. In the figure \ref{results}, we list all diagrams giving a non-zero contribution to the three-point functions as well as the result of the respective scheme independent constants. 
A relevant aspect of this computation is that some terms in the second line of (\ref{eq:rotation}) are now important at one-loop level. Indeed, from figure \ref{results} we realize that the second graph of the second row mixes up the $R$-charge indices of the scalar and the fermion. In particular, the scalar $\Phi^{14}$ and the fermion $\psi^2$ in the second line of  (\ref{eq:rotation}) can be converted into a $\Psi$ and a $\bar{Z}$ through this diagram. The resulting state can then be contracted with the remaining external operators and give a non-vanishing contribution.
\\
\indent From the results of figure \ref{results}, we can directly read off an operator acting on the two fields at the splitting points of an external state and that gives those same constants after contraction with the remaining states. We denote this operator by $\mathcal{F}$ and define it by the following matrix elements
\begin{alignat}{3}
&\langle \, \psi^{a} \, \psi^{b} \, | \, \mathcal{F} \, | \, \psi^{c} \, \psi^{d} \, \rangle = - \,  \delta^{ac}\delta^{bd} \, ,  \label{Foperator} \\ & \nonumber \\
& \langle \, \Phi^{ef} \, \Phi^{gh} \, | \, \mathcal{F}\,  | \, \Phi^{ab} \, \Phi^{cd} \, \rangle = 2 \, \bar{\delta}^{gh,ab} \,  \bar{\delta}^{ef,cd} - 2 &&\, \bar{\delta}^{ef,ab} \, \bar{\delta}^{gh,cd} - \, \epsilon^{abcd} \, \epsilon^{efgh} \, ,   \nonumber \\ & \nonumber  \\
& \langle \, \Phi^{de} \, \psi^f \, | \, \mathcal{F} \, | \, \Phi^{ab} \, \psi^c \, \rangle = - \, \delta^{fc}\bar\delta^{ab,de}  \,,  &&\langle \, \psi^f \, \Phi^{de} \,  | \, \mathcal{F} \, | \, \psi^c \, \Phi^{ab} \, \rangle = - \, \delta^{fc}\bar\delta^{ab,de} \, , \nonumber
\\ & \nonumber \\
& \langle \, \Phi^{de} \, \psi^f \, | \, \mathcal{F} \, | \, \psi^c \, \Phi^{ab} \,  \rangle =  \delta^{ce}\bar\delta^{ab,df}\,, &&\langle \, \psi^f \, \Phi^{de} \,  | \, \mathcal{F} \, | \, \Phi^{ab} \, \psi^c \,  \rangle =  \delta^{ce}\bar\delta^{ab,df} \, , \nonumber
\end{alignat}
where $\bar\delta^{ab,cd}\equiv\delta^{ac}
\delta^{bd}-\delta^
{ad}\delta^{bc}$ and in the second line 
we recognize the $\alg{so}(6)$ Hamiltonian \cite{Zarembo, Alday, Japas}. 
It is simple to check that the operator $\frac{g^2}{2}\mathcal{F}$ reproduces the constants of figure \ref{results}.

For the specific setup that we are considering only the diagrams of figure
\ref{results} are relevant, since additional diagrams either cancel among them or vanish, 
see Appendix 
\ref{appdetails} for details. In the case of a more general setup, the operator $\mathcal{F}$ defined receives corrections from new diagrams.  

In what follows, the operator $\mathcal{F}$ will appear with additional indices as $\mathcal{F}_{ij}$, which indicate the sites in the spin chain where the operator acts. As an example, we have that
\begin{eqnarray}
\langle \, \ldots \stackrel{i}{\Psi} \stackrel{j}{Z} \ldots \, | \, \frac{g^2}{2}\mathcal{F}_{ij} \, |\,
\ldots \stackrel{i}{\Psi} \stackrel{j}{Z} \ldots \, \rangle = - \frac{g^2}{2} \, , \nonumber   
\end{eqnarray}
which reproduces the result of the first diagram of the second row of figure \ref{results}. It is 
important to note that when the operator 
$\mathcal{F}_{ij}$ acts on non-neighboring sites, it can pick up additional minus signs
due to statistics, for example, 
\begin{eqnarray}
\langle \, \Psi \ldots \underbrace{{\Psi} \ldots {\Psi}}_{n\text{ fermions}}  \ldots Z \, | \, \frac{g^2}{2}\mathcal{F}_{1L} \, |\,
Z \ldots \underbrace{{\Psi} \ldots {\Psi}}_{n\text{ fermions}} \ldots \Psi \, \rangle = (-1)^{n} \, \frac{g^2}{2} \, , \nonumber
\end{eqnarray}
where $n$ denotes the number of fermionic excitations between the first and last sites and 
we have used the last rule of (\ref{Foperator}). 
\subsection{Final result}
We now give the complete expression for the structure constants up to one-loop in the setup considered in this work. It reads
\<
C_{123}=\alpha\times \Bigl(  &&\langle \bold{1}^f |1+ \frac{g^2}{2}\mathcal{F}_{L_3-N_3,L_3-N_3+1}  +\frac{g^2}{2} \mathcal{F}_{L_{1}, 1} |\underbrace{\bar{Z}\dots\bar{Z}}_{L_3-N_3} i_1 \ldots i_{L_2-N_3} \rangle   \label{final}\\
&&  \langle  \underbrace{\bar\Psi\dots \bar\Psi}_{N_3} i_1 \ldots i_{L_2-N_3}  |1+ \frac{g^2}{2} \mathcal{F}_{N_{3},N_{3}+1} +\frac{g^2}{2}\mathcal{F}_{L_{2}, 1}  | \bold{2} \rangle   \Bigr) \times \nonumber \\
&& \langle\underbrace{\Psi\dots\Psi }_{N_3} \underbrace{ \bar{Z}\dots\bar{Z}}_{L_3-N_3}  | 1 +\frac{g^2}{2} \mathcal{F}_{ N_3,N_3 +1} +\frac{g^2}{2}\mathcal{F}_{L_{3}, 1}  | \bold{3} \rangle\, \nonumber,
\>where we have that

\[
\alpha=\sqrt{\frac{L_1 L_2 L_3}{\mathcal{N}^{(1)} \mathcal{N}^{(2)} \mathcal{N}^{(3)}}} \, ,\label{alpha}
\]
with $\mathcal{N}^{(i)}$ being the respective norms and we are using the conventions
\begin{equation}
\langle \sigma_{i_1} \sigma_{i_2} \cdots \sigma_{i_L}|\sigma_{j_1} \sigma_{j_2} \cdots \sigma_{j_L}\rangle = \delta_{i_1j_1}\delta_{i_2j_2}
\cdots\delta_{i_Lj_L} \, , \nonumber
\end{equation} 
where $\sigma$ is any field.

In the formula (\ref{final}), $i_a$ can be either $\bar{Z}$ or $\bar\Psi$ and a sum over all these intermediate states is implied. Moreover, we have included a superscript $f$ in the bra associated to the operator
$\mathcal{O}_1$ to emphasize that the state was \emph{flipped}\footnote{In short, the \emph{flipping} operation $\mathcal{F}l$ introduced in \cite{TailoringI} is defined as $\mathcal{F}l:\,\,\, \psi(n_1,\dots,n_N)|n_1,\dots,n_N\rangle \mapsto \psi(n_1,\dots,n_N)\langle L-n_N+1,\dots, L-n_1+1| \,\hat{C}$, where $\hat{C}$ means charge conjugation which exchanges $Z \leftrightarrow \bar{Z}$ and $\Psi \leftrightarrow \bar{\Psi}$.}, see \cite{TailoringI} for details. The external states are the two-loop corrected Bethe eigenstates as described in section \ref{two loop}, for instance
\[
|\bold{1}\rangle = |1\rangle^{(0)}+g^2 |1\rangle ^{(1)}+\mathcal{O}(g^4)\,.
\]

We have checked that for the simple case of three half-BPS operators, the one-loop correction to the structure constant vanishes as expected from the non-renormalization theorem of \cite{Minwalla}, see Appendix \ref{examples} for details. Additionally, in the Appendix \ref{invariance} we check that this result satisfies some constraints from symmetry considerations.\\
\indent 

The expression (\ref{final}) can now be evaluated as an explicit function of the Bethe roots by using the known form of the two-loop Bethe states. As the number of excitations on the external states increases, such task becomes tedious and the result gets lengthy obscuring possible simplifications. Nevertheless, we can easily deal with states of arbitrary length but only a few magnons. It turns out that the manipulation of the resulting expressions for these simple cases reveals a strikingly compact structure that can be easily generalizable for arbitrary complicated states. We then resort to the numerical approach in order to confirm that such generalization actually holds. In the end, we find a formula given by a very simple and natural deformation of the tree-level result (\ref{finaltree}), as follows
\<
C_{123}=\mathcal{C}\;\frac{\prod\limits_{k=1}^{3}\prod\limits_{\substack{i< j}}^{N_k}\mathfrak{f}(y^{(k)}_i,y^{(k)}_j)}{\prod\limits_{i=1}^{N_1}\prod\limits_{j=1}^{N_2}\mathfrak{f}(y^{(1)}_i,y^{(2)}_j)}\prod\limits _{k=1}^{N_1} \left[1-(y^{(1)}_{k})^{L_2}\prod\limits_{i=1}^{N_2} \left(-S(y^{(2)}_i,y^{(1)}_k)\right)\right]\,, \label{oneloop}
\>
where we are using the notation $y^{(i)}_k=e^{i p^{(i)}_k}$ and the normalization factor $\mathcal{C}$ is given by
\[
\mathcal{C}=\sqrt{\frac{L_1 L_2 L_3}{ \mathcal{N}^{(1)} \mathcal{N}^{(2)} \mathcal{N}^{(3)}}}\left[1+g^2 \left(N_3^2-1\right)-\frac{1}{4}\sum_{i=1}^{3}\bold\gamma_i \right]\,, \label{normalfactor}
\]
with $\gamma_i$ being the anomalous dimension of the operator $\mathcal{O}_i$. As described in the section \ref{two loop}, the norms $ \mathcal{N}^{(i)}$ are given by the formula
\[
\mathcal{N}^{(i)} =   \det_{j,k\le N_i} \frac{\partial }{\partial p^{(i)}_j} \left[ L p^{(i)}_{k}
+\frac{1}{i}\sum_{m\neq k}^{N_i} \log S(p^{(i)}_m,p^{(i)}_k) \right]\,.
\]
The most important and non-trivial part of the final result is the function $\alg{f}$ which reads
\[
\mathfrak{f}(s,t)=(s-t)\left[1-\frac{g^2}{2}\left(\frac{s}{t}+\frac{t}{s}-\frac{1}{s}-s-\frac{1}{t}- t +2\right)\right]\,.
\]
The momenta  $p^{(j)}_{k}$ of the fermionic excitations
must satisfy the Bethe equations which take the form
\[
e^{i p^{(j)}_k L_j}=\prod_{i\neq k}^{N}\left( - S(p^{(j)}_k,p^{(j)}_i) \right)\, ,  \label{BAE2loop}
\]
and the total momentum condition (\ref{totalm}).
This constitutes the most important result of this paper and it will be discussed in the next section.

\section{Discussion and open problems}

In this work, we have computed both the leading order contribution and the one-loop perturbative correction at weak coupling
to the three-point functions of single-trace operators of ${\mathcal{N}}=4$ SYM in the $\alg{su}(1|1)$ sector. The $\alg{su}(1|1)$ sector is closed to all orders in perturbation theory \cite{BeisertTwoloop} and it is the simplest sector having both fermions
and bosons. 
Representing each operator by a Bethe eigenstate, we were able to derive a simple expression (\ref{finaltree}) for the leading order result in  terms of the momenta characterizing the states.\\
\indent In addition, we have also computed the one-loop correction by evaluating the relevant Feynman diagrams and also determining the two-loop Bethe eigenstates and their norm. The prescription for computing the scheme independent three-point structure constant turns out to be given in terms of the insertion of the operator $\mathcal{F}$, defined in (\ref{Foperator}), at the splitting points of the spin chains. \\
\indent Regarding the external states, due to the long-range property of the two-loop dilatation operator in the $\alg{su}(1|1)$ sector, its diagonalization involves the usual Bethe ansatz corrected by the contact terms. These in turn are independent of the momenta of the excitations and have a very simple expression for an arbitrary number of magnons, see (\ref{cterms}). The norm of these states is compactly given by a simple determinant, analogous to the well known case of the $\alg{su}(2)$ sector.\\
\indent The one-loop structure constant in our setup turns out to be given by the simple formula 
(\ref{oneloop})
in terms of the Bethe roots. It is tempting to investigate the thermodynamic limit of our result, namely when we consider one or more long spin chains $L_i\gg 1$, with a large number of excitations $N_i=\mathcal{O}(L_i)$. This might be useful for future comparison with string theory calculations in a specific limit. An obvious open problem is the computation of the three-point functions in higher rank sectors at least at tree-level. Our result and the results of \cite{FodaMetodo,Pedrosl2} are encouraging in order to find a simple expression for the full $\alg{psu}(2,2|4)$. The main obstacle is the knowledge of the scalar products of Bethe states for generic (super) algebras, although some progress has been made in the $\alg{su}(3)$ case \cite{Wheelersu3,WheelerNovo,Slavnov1,Slavnov2}.  

\indent  The final expression 
(\ref{oneloop}) is very suggestive and deserves further comments.
Apart from the simple normalization factor $\mathcal{C}$ given by the expression (\ref{normalfactor}), the structure constant has two distinct contributions. Firstly, the one-loop correction to the $S$-matrix appears in a very natural way when we look at the tree-level result (\ref{finaltree}). Secondly, the most non-trivial part comes from the function $\alg{f}$. The one-loop result is achieved by deforming this function, which bears some similarities to the $\alg{su}(2)$ and $\alg{sl}(2)$ cases \cite{TailoringIV,Pedrosl2}. As already pointed out in \cite{Pedrosl2}, it would be interesting to deepen the connection of the three-point function with form factors as started in \cite{formfactors1,formfactors2}. In particular, that could shed light on a (non-perturbative) definition of the function $\alg{f}$ from the form factors axioms. In fact, such axiomatic approach was recently explored in the context of the scattering amplitudes \cite{fluxtube1,fluxtube2,fluxtube3}. There, the central object called pentagon transition $P(u|v)$ was required to satisfy some natural constraints from the integrability point of view. These conditions were then used to bootstrap the function exactly. 
 In this regard, we notice some striking similarities of the dependence of our final result on this function $\alg{f}$ with the expression (9) of \cite{fluxtube1} which  corresponds to a multi-particle transition.
We hope that such ideas can be applied for the calculation of three-point functions at any value of the coupling constant.

\section*{Acknowledgements}
We would like to thank P. Vieira for many invaluable discussions and suggestions. We also thank B. Basso, N. Berkovits, N. Gromov, Y. Jiang, H. Nastase, J. Penedones, D. Serban, A. Sever, C. Sieg, E. Sobko,  J. Toledo for comments and  discussions and especially T. Wang for collaboration in a stage of this project. We would like to thank the warm hospitality of the ICTP-SAIFR, FAPESP grant 2011/11973-4, where part of this work was done. TF would like to thank the warm hospitality of the Perimeter Institute where this work was initiated.\\
JC is funded by the FCT fellowship SFRH/BD/69084/2010. This work has been supported in part by the Province of Ontario through ERA grant ER 06-02-293. Research at the Perimeter Institute is supported in part by the Government of Canada through NSERC and by the Province of Ontario through MRI. This work was partially funded by the research grants PTDC/FIS/099293/2008 and CERN/FP/116358/2010. \emph{Centro de F\'{i}sica do Porto} is partially funded by FCT under grant PEst-OE/FIS/UI0044/2011. 
TF would
like to thank FAPESP grant
2013/12416-7 and 2009/50775-3 for financial support.

\appendix
\section{Notation and conventions}\label{appnotations}
In this Appendix, we fix our conventions for the perturbative computations. The $\mathcal{N}=4$ SYM with $SU(N)$ gauge group has the following Lagrangian \cite{Georgiou:2008vk,Georgiou:2012zj} 
\<\mathcal{L}=\text{Tr}\Bigl(&& -\frac{1}{2}F_{\mu \nu}F^{\mu \nu}+2 \mathcal{D}_{\mu}\Phi_{ab}\,\mathcal{D}^{\mu}\Phi^{ab}+2i\psi^{\alpha a}\sigma_{\alpha \dot{\alpha}}^{\mu} (\mathcal{D}_{\mu} \bar{\psi}_a)^{\dot{\alpha}}  \\
&&+ 2 g^{2}_{\text{YM}} [\Phi^{ab},\Phi^{cd}][\Phi_{ab},\Phi_{cd}] -2 \sqrt{2} g_{\text{YM}}\left([\psi^{\alpha a},\Phi_{ab}]\psi^{b}_{\alpha}-[\bar{\psi}_{\dot{\alpha}a},\Phi^{ab}]\bar{\psi}_{b}^{\dot{\alpha}}\right)\Bigr)\,, \nonumber
\> 
with all the fields in the adjoint representation of the gauge group and the covariant derivative is $\mathcal{D}_{\mu} \,\cdot=\partial_{\mu}- i g_{YM} [A_{\mu},\,\cdot\,]$. The propagators extracted from this Lagrangian are (we are suppressing the gauge indices and taking the leading order in $N$)
\<
&&\langle \Phi^{ab}(x)\Phi_{cd}(0) \rangle = \frac{\bar\delta_{cd}^{ab}  }{8}  \frac{1}{(2\pi)^2(-x^2+i \epsilon)} \, , \nonumber \\ 
&&\langle \psi^{a}_{\alpha}(x)\bar\psi_{\dot{\beta}\, b}(0) \rangle =\frac{i\, \delta^{a}_{b}}{2} \sigma^{\mu}_{\alpha \dot{\beta}}\partial_{\mu}\frac{1}{(2\pi)^2(-x^2+i \epsilon)} \, , \nonumber \\ 
&&\langle A_{\mu}(x) A_{\nu}(0) \rangle = -\frac{\eta_{\mu \nu}}{2}\frac{1}{(2\pi)^2(-x^2+i \epsilon)} \, , \nonumber 
\>
where $\bar\delta_{cd}^{ab}\equiv\delta_{c}^{a}\delta_{d}^{b}-\delta_{c}^{b}\delta_{d}^{a}$. We are using the Minkowski metric $(+-\,-\;-)$ and the Feynman gauge. The action of the (classical) supersymmetry generators are given by \cite{Korchemsky}
\<
&&[\alg{Q}^{\alpha}_{\;\;a}, \Phi^{bc}]= \frac{i\sqrt{2}}{2}\left(\delta^{b}_a \psi^{\alpha c}-\delta^{c}_a \psi^{\alpha b}\right) \, ,
 \nonumber \\ 
&&[\alg{Q}^{\alpha}_{\;\;a}, \psi^{b}_{\beta}]= \delta_{a}^{b}F^{\alpha}_{\beta} \, , \nonumber \\ 
&&[\alg{Q}^{\alpha}_{\;\;a}, \bar{\psi}_{b}^{\dot{\beta}}]= 2\sqrt{2}\,\mathcal{D}^{\dot{\beta} \alpha}\Phi_{ab}\, , \nonumber
\>
and the conjugate expressions for the action of $\alg{\bar Q}^{a}_{\;\;\dot{\alpha}}$\,. The action of the 
$R$-symmetry generators is given by
\<
&&[\alg{R}^{a}_{\;\;b}, \Phi^{cd}]=\delta^{c}_b\, \Phi^{ad}+\delta^{d}_b\, \Phi^{ca}-\frac{1}{2}\delta^{a}_b \, \Phi^{cd} \, ,
 \nonumber \\ 
&&[\alg{R}^{a}_{\;\;b}, \psi^{c}]=\delta^{c}_b\, \psi^{a}-\frac{1}{4}\delta^{a}_b\,  \psi^{c} \, .
 \nonumber 
\>

In the computations of the Feynman diagrams, in particular for the evaluation of the integrals, we analytical continued to Euclidean space by using  
\begin{equation}
x^{0} = i x^4 \, , \quad \quad \quad \sigma^{0}_M = - \sigma^0_E \, , \quad \quad \quad \sigma^i_M = i \sigma^i_E \, \nonumber,
\end{equation}
where the subscripts $M$ means Minkowski space and $E$ means Euclidean space, 
\begin{equation}
\sigma^0_M =\text{Id}_{2\times 2} \, ,  \nonumber
\end{equation} 
and, finally, $\sigma_M^i$ are the usual Pauli matrices.

\section{One-loop perturbative computation details}\label{appdetails}

\begin{figure}[t]
\begin{center}
\includegraphics[width=1.0\linewidth]{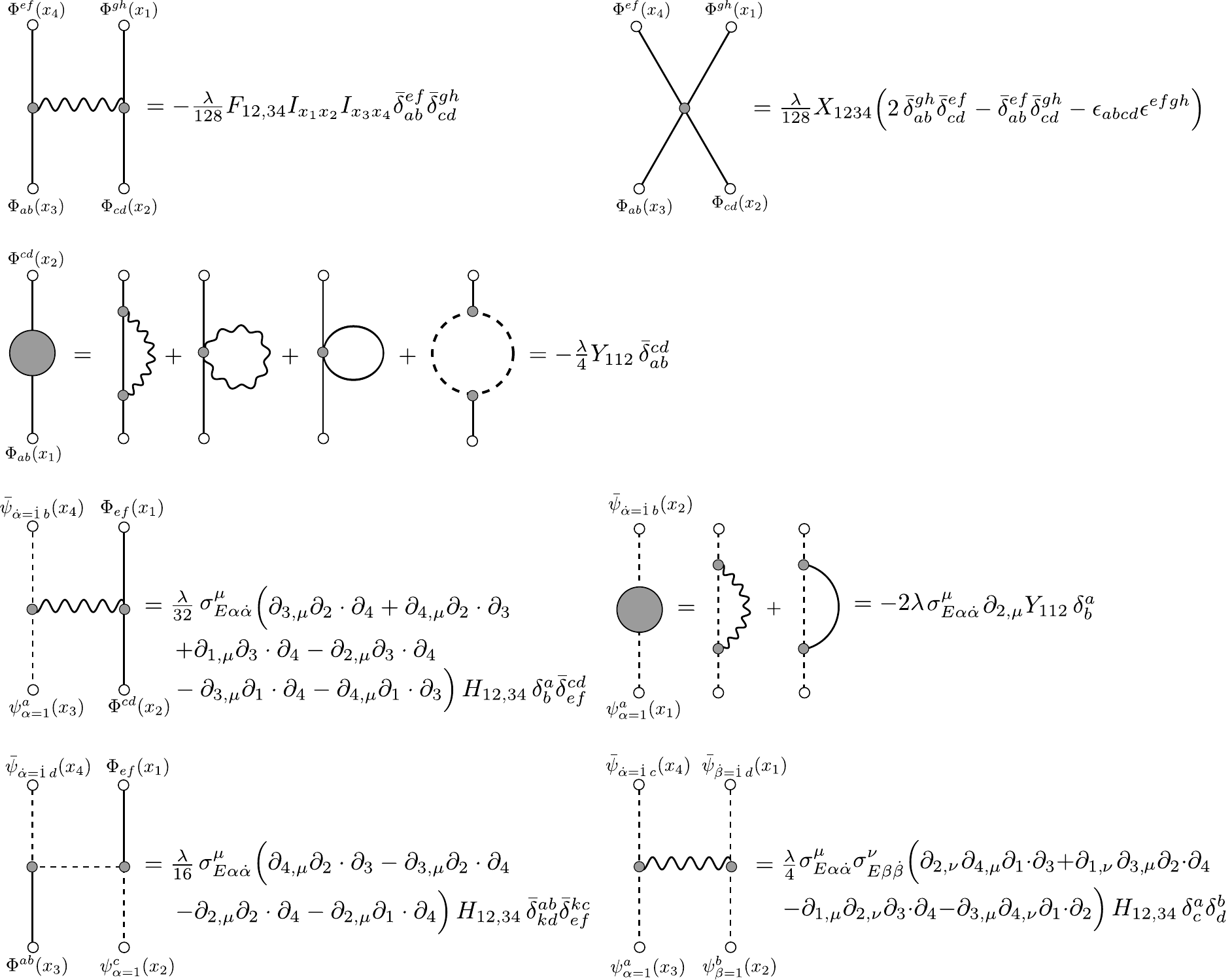}
\caption{The results of the Feynman diagrams computation omitting both terms that must vanish or cancel when summing all the diagrams (see text) and factors of $N$. 
The solid, wiggly and dashed lines represent the scalars, gluons and fermions, respectively. 
The $\bar{\delta}$ was defined in Appendix A. \label{fulldetails} }
\end{center}
\end{figure}

In this Appendix, we present the details of the perturbative computation of the three-point functions at one-loop using the point splitting regularization. As reviewed in the main part of this paper, in order to obtain scheme and normalization independent structure constants we also need to know the results of the two-point functions. For completeness, we explicitly compute the one-loop dilatation operator of the $\alg{su}(1|1)$ sector as well. 
\\
\indent Typically three kinds of integrals will appear in the computations
\<
&&Y_{123}=\int d^{4}u \,I_{x_1 u} I_{x_2 u} I_{x_3 u} \, , \nonumber \\
&&X_{1234} = \int d^{4}u \,I_{x_1 u} I_{x_2 u} I_{x_3 u} I_{x_4 u} \, , \nonumber \\
&&H_{12, 34} = \int d^{4}u\, d^{4}v\, I_{x_1 u} I_{x_2 u} I_{u v} I_{x_3 v} I_{x_4 v} \, , \nonumber
\>
where $I_{x_a x_b}$ is the (euclidean) scalar propagator defined as
\[
I_{x_a x_b}\equiv\frac{1}{(2 \pi)^2 (x_{a}-x_{b})^2}\,. \nonumber
\]
The $Y$ and $X$ integrals are well-known and explicit expressions for them can be found for instance in \cite{BeisertBMN, Alday}. The integral $H$ is not known analytic, however, only its derivatives will be needed. In particular, the following combination \cite{BeisertBMN} turns out to be useful 
\<
F_{12,34}\,&\equiv&\, \frac{(\partial_1-\partial_2)\cdot(\partial_3 -\partial_4) H_{12,34}}{I_{x_1 x_2} I_{x_3 x_4}} \label{F}\\
&=&\,\frac{X_{1234}}{I_{x_1 x_3}I_{x_2 x_4}}-\frac{X_{1234}}{I_{x_1 x_4}I_{x_2 x_3}}+G_{1,34}-G_{2,34}+G_{3,12}-G_{4,12}\, , \nonumber
\>where 
\[
G_{a,bc}=\frac{Y_{abc}}{I_{x_a x_c}}-\frac{Y_{abc}}{I_{x_a x_b}}\,. \nonumber
\]
\indent We will need several limits of the expressions for $Y$ and $X$, namely when pairs of distinct points collapse into each other
\<
Y_{113}&\,\equiv\,& \lim_{x_2\rightarrow x_1} Y_{123}\,=\,   \left(2-\log\left(\frac{\epsilon ^2}{x_{13}^2}\right)\right)\frac{I_{x_1 x_3}}{16\pi^2} \, ,        \nonumber \\
X_{1134}&\,\equiv\,& \lim_{x_2\rightarrow x_1} X_{1234}\,=\,\left(2-\log\left(\frac{\epsilon ^2 x_{34}^2}{x_{13}^2 x_{14}^2}\right)\right)\frac{I_{x_1 x_3} I_{x_1 x_4}}{16\pi^2} \, , \nonumber
\>
where we are considering $x_2^{\mu}=x_1^{\mu}+\epsilon^{\mu}$ with $\epsilon^{\mu}\rightarrow 0$.
We can also take a further limit of the last expression above when $x_4\rightarrow x_3$ giving
\[
X_{1133}=\left(1-\log\left(\frac{\epsilon ^2}{x_{13}^2}\right)\right)\frac{I_{x_1 x_3}^2}{8 \pi ^2}\,. \nonumber
\]
Moreover, we also need limits of the first and second derivatives of both the $Y$ and the $X$ integrals. We include the results of them below for completeness. The first derivatives are given by
\<
\lim_{x_2\rightarrow x_1}\partial_{1,\mu}Y_{123}\;&=&\;\frac{\epsilon_{\mu} }{ \epsilon ^2}\frac{I_{x_1 x_3}}{8 \pi ^2}- \left(1-\log \left(\frac{\epsilon ^2}{x_{13}^2}\right)\right) \frac{x_{13,\mu} I_{x_1 x_3}^2 }{4}- \frac{x_{13,\nu} \epsilon^{\nu} \epsilon_{\mu} I^2_{x_1x_3}}{2 \epsilon^2}\, ,\label{oneovereps}\\
\lim_{x_2\rightarrow x_1}\partial_{3,\mu}Y_{123}\;&=&\; \left(1-\log \left(\frac{\epsilon ^2}{x_{13}^2}\right)\right)\frac{  x_{13,\mu }  I_{x_1 x_3}^2}{2} \, , \label{dY}\\
\lim_{x_2\rightarrow x_1}\partial_{1,\mu}X_{1234}\;&=&\; \frac{ \epsilon_{\mu}}{ \epsilon ^2}\frac{I_{x_1 x_3} I_{x_1 x_4}}{8 \pi ^2}-\left(1- \log\left(\frac{\epsilon ^2 x_{34}^2}{x_{13}^2 x_{14}^2}\right)\right) \frac{x_{13,\mu } I_{x_1 x_3}^2 I_{x_1 x_4} + x_{14,\mu} I_{x_1 x_3} I_{x_1 x_4}^2 }{4} \nonumber \\
& & -\frac{x_{14,\nu} \, \epsilon^{\nu} \epsilon_{\mu} I_{x_1 x_3} I_{x_1 x_4}^2}{2 \epsilon^2} -\frac{x_{13,\nu}\, \epsilon^{\nu} \epsilon_{\mu} I_{x_1 x_3}^2 I_{x_1 x_4}}{2 \epsilon^2} \, ,  \nonumber \\
\lim_{x_2\rightarrow x_1}\partial_{3,\mu}X_{1234}\;&=&\;-\frac{x_{34,\mu }I_{x_1 x_3} I_{x_1 x_4} I_{x_3 x_4}}{2} + \left(1-\log\left(\frac{\epsilon ^2 x_{34}^2}{x_{13}^2 x_{14}^2}\right)\right) \frac{x_{13,\mu} I_{x_1 x_3}^2 I_{x_1 x_4} }{2} \, , \nonumber
\>
As before, one can take further limits of these expressions when needed. The second derivatives read
\<
\lim_{x_2\rightarrow x_1}\partial_{1,\mu}\partial_{2,\nu}Y_{123}&\;=\;&-\frac{\epsilon_{\mu}\epsilon_{\nu}}{\epsilon ^4}\frac{I_{x_1 x_3} }{4 \pi ^2 }+ \frac{\epsilon_{\mu} \epsilon_{\nu} I_{x_1 x_3}^2}{\epsilon^2}\, ( \,  \frac{1}{6}+ \frac{x_{13,\rho} \, \epsilon^{\rho}}{\epsilon^2}- \frac{16 \pi^2 x_{13, \rho} \, \epsilon^{\rho} x_{13, \sigma} \, \epsilon^{\sigma}  I_{x_1x_3}}{3 \epsilon^2} \, ) \nonumber \\ & +& \frac{ \epsilon_{\nu}}{\epsilon ^2}\frac{x_{13,\mu }I_{x_1 x_3}^2 }{2}-\frac{\epsilon_{\mu}}{\epsilon ^2}\frac{x_{13,\nu }I_{x_1 x_3}^2  }{2 }+ \frac{8 \pi^2 x_{13,\rho} \, \epsilon^{\rho} I_{x_1 x_3}^3}{3 \epsilon^2} \, (2 x_{13,\nu} \epsilon_{\mu} - x_{13, \mu} \epsilon_{\nu})   \nonumber \\ &+& \frac{1}{ \epsilon ^2}\frac{ \delta_{\mu \nu }I_{x_1 x_3}}{8 \pi ^2}- \delta_{\mu \nu} I_{x_1 x_3}^2 \, ( \, \frac{11}{36} + \frac{x_{13,\rho} \epsilon^{\rho} }{2 \epsilon^2} - \frac{8 \pi^2 x_{13, \rho} \, \epsilon^{\rho} x_{13, \sigma} \, \epsilon^{\sigma}I_{x_1x_3}}{3 \epsilon^2} ) \nonumber\\
&+&\frac{1}{12} \log\left(\frac{\epsilon ^2}{x_{13}^2}\right) I_{x_1 x_3}^2 \delta_{\mu \nu}
+\frac{2\pi ^2 }{9} \left(1-6 \log\left(\frac{\epsilon ^2}{x_{13}^2}\right)\right)x_{13,\mu}x_{13,\nu} I_{x_1 x_3}^3  \, ,\nonumber \\
\lim_{x_2\rightarrow x_1}\partial_{1,\mu}\partial_{3,\nu}Y_{123}&\;=\;&\frac{\epsilon_{\mu} \epsilon_{\nu}}{\epsilon ^2}\frac{I_{x_1 x_3}^2 }{2 }+\frac{\epsilon_{\mu}}{\epsilon ^2} x_{13,\nu }I_{x_1 x_3}^2-2 \pi ^2 \left(1-2 \log\left(\frac{\epsilon ^2}{x_{13}^2}\right)\right)x_{13,\mu } x_{13,\nu } I_{x_1 x_3}^3 \nonumber \\
&+&\frac{1}{4} \left(1-\log\left(\frac{\epsilon ^2}{x_{13}^2}\right)\right) I_{x_1 x_3}^2 \delta_{ \mu \nu } - \frac{8 \pi^2 x_{13, \nu} \, \epsilon_{\mu}\,  x_{13, \rho} \, \epsilon^{\rho} I_{x_1 x_3}^3 }{\epsilon^2} \, ,  \label{lastY}
\>
\<
\lim_{x_2\rightarrow x_1}\partial_{1,\mu}\partial_{2,\nu}X_{1234}&\;=\;& -\frac{ \epsilon_{\mu} \, \epsilon_{\nu}}{\epsilon ^4}\frac{I_{x_1 x_3} I_{x_1 x_4}}{4 \pi ^2 } + \frac{\epsilon_{\mu} \, \epsilon_{\nu} I_{x_1 x_3}^2 I_{x_1 x_4}^2}{6 \epsilon^2}\, ( \, \frac{1}{I_{x_3 x_4}}- \frac{32 \pi^2 x_{13,\rho} \, \epsilon^{\rho} \, x_{14, \sigma} \, \epsilon^{\sigma}}{ \epsilon^2})\nonumber\\
&+& \frac{\epsilon_{\mu} \, \epsilon_{\nu} \, \epsilon^{\rho} I_{x_1 x_3} I_{x_1 x_4}}{\epsilon^4}\, (\, x_{14, \rho} I_{x_1x_4}+ x_{13, \rho} I_{x_1x_3})\nonumber \\
&-& \frac{16 \pi^2 \, \epsilon_{\mu} \, \epsilon_{\nu} \, \epsilon^{\rho} \, \epsilon^{\sigma} I_{x_1x_3}I_{x_1x_4}}{3 \epsilon^4} \, ( \, x_{14, \rho} x_{14, \sigma} I_{x_1 x_4}^2 + x_{13, \rho} x_{13, \sigma} I_{x_1 x_3}^2) \nonumber \\
&-& \frac{\epsilon_{\mu} I_{x_1x_3} I_{x_1 x_4}}{2 \epsilon^2} \, (\, x_{14, \nu} I_{x_1 x_4} + x_{13,\nu} I_{x_1 x_3}) \nonumber \\ &+& \frac{\epsilon_{\nu} I_{x_1x_3} I_{x_1 x_4}}{2 \epsilon^2} \, (\, x_{14, \mu} I_{x_1 x_4} + x_{13,\mu} I_{x_1 x_3}) \nonumber \\
&+& \frac{8 \pi^2 \epsilon_{\mu} \, \epsilon^{\rho} I_{x_1x_3}I_{x_1x_4}}{3 \epsilon^2} \, (\, 2 x_{14,\nu} \, x_{14,\rho} \,  I_{x_1x_4}^2+x_{13,\rho} \, x_{14,\nu} \, I_{x_1x_3}I_{x_1x_4}) \nonumber \\
&+& \frac{8 \pi^2 \epsilon_{\mu} \, \epsilon^{\rho} I_{x_1x_3}I_{x_1x_4}}{3 \epsilon^2} \, (\, 2 x_{13,\nu} \, x_{13,\rho} \,  I_{x_1x_3}^2+x_{14,\rho} \, x_{13,\nu}\, I_{x_1x_3}I_{x_1x_4}) \nonumber \\
&-& \frac{4 \pi^2 \epsilon_{\nu} \, \epsilon^{\rho} I_{x_1x_3}I_{x_1x_4}}{3 \epsilon^2} \, (\, 2 x_{14,\mu} \, x_{14,\rho} I_{x_1x_4}^2+x_{13,\rho} \, x_{14,\mu}I_{x_1x_3}I_{x_1x_4}) \nonumber \\
&-& \frac{4 \pi^2 \epsilon_{\nu} \, \epsilon^{\rho} I_{x_1x_3}I_{x_1x_4}}{3 \epsilon^2} \, (\, 2 x_{13,\mu} \, x_{13,\rho} I_{x_1x_3}^2+x_{14,\rho} \, x_{13,\mu}I_{x_1x_3}I_{x_1x_4}) \nonumber \\
&+& \frac{1}{\epsilon ^2}\frac{I_{x_1 x_3} I_{x_1 x_4} \delta_{\mu \nu }}{8 \pi ^2 } - \frac{I_{x_1x_3}I_{x_1x_4} \delta_{\mu \nu}}{4} \, (\, I_{x_1x_4}+I_{x_1x_3})\nonumber\\
&-& \frac{ I_{x_1 x_3} I_{x_1 x_4} \delta_{\mu \nu} \, \epsilon^{\rho}}{2 \epsilon^2} \, (\, x_{14, \rho} \, I_{x_1 x_4} + x_{13, \rho} \, I_{x_1x_3}) \nonumber \\
&+& \frac{8 \pi^2 I_{x_1x_3} I_{x_1x_4} \, \delta_{\mu \nu} \, \epsilon^{\rho} \, \epsilon^{\sigma}}{3 \epsilon^2} \, ( \, x_{14,\rho} \, x_{14, \sigma} \, I_{x_1 x_4}^2 + x_{13, \rho} \, x_{13, \sigma} \, I^2_{x_1 x_3}) \nonumber \\
&+& \frac{8 \pi^2 I_{x_1x_3} I_{x_1x_4} \, \delta_{\mu \nu} \, \epsilon^{\rho} \, \epsilon^{\sigma}}{3 \epsilon^2} \, x_{13, \rho} \, x_{14, \sigma} \, I_{x_1 x_3} I_{x_1 x_4} \nonumber \\
&+&\frac{1}{36}\left(-2+3 \log\left(\frac{\epsilon ^2 x_{34}^2}{x_{13}^2 x_{14}^2}\right)\right)\frac{ I_{x_1 x_3}^2 I_{x_1 x_4}^2 \delta_{\mu \nu}}{I_{x_3 x_4}}  \nonumber \\
 &+&\frac{2 \pi ^2}{9} \left(1-6 \log\left(\frac{\epsilon ^2 x_{34}^2}{x_{13}^2 x_{14}^2}\right)\right)x_{13,\mu } x_{13,\nu } I_{x_1 x_3}^3 I_{x_1 x_4} \nonumber\\
&-&\frac{2 \pi ^2}{9} \left(1+3 \log\left(\frac{\epsilon ^2 x_{34}^2}{x_{13}^2 x_{14}^2}\right)\right)\left(x_{13,\nu } x_{14,\mu }+ x_{13,\mu } x_{14,\nu }\right) I_{x_1 x_3}^2 I_{x_1 x_4}^2 \nonumber\\
&+&\frac{2 \pi ^2}{9} \left(1-6 \log\left(\frac{\epsilon ^2 x_{34}^2}{x_{13}^2 x_{14}^2}\right)\right) x_{14,\mu } x_{14,\nu }I_{x_1 x_3} I_{x_1 x_4}^3 \, , \nonumber\\
\lim_{x_2\rightarrow x_1}\partial_{1,\mu}\partial_{3,\nu}X_{1234}&\;=\;&\frac{\epsilon_{\mu} \epsilon_{\nu}}{ \epsilon ^2}\frac{I_{x_1 x_3}^2 I_{x_1 x_4} }{2}+2 \pi ^2 \log\left(\frac{\epsilon ^2 x_{34}^2}{x_{13}^2 x_{14}^2}\right)x_{13,\nu } x_{14,\mu } I_{x_1 x_3}^2 I_{x_1 x_4}^2\nonumber\\
&+&\frac{x_{13, \nu} \, \epsilon_{\mu} I^2_{x_1x_3}I_{x_1x_4}}{\epsilon^2} \, (\, 1 - 4 \pi^2 x_{14,\rho} \, \epsilon^{\rho} I_{x_1x_4} - 8 \pi^2 x_{13, \rho} \, \epsilon^{\rho} I_{x_1x_3})\nonumber \\
&+&2 \pi ^2 \left(-1+2 \log\left(\frac{\epsilon ^2 x_{34}^2}{x_{13}^2 x_{14}^2}\right)\right)x_{13,\mu } x_{13,\nu } I_{x_1 x_3}^3 I_{x_1 x_4} \nonumber\\
&-&\frac{1}{4} \left(-1+\log\left(\frac{\epsilon ^2 x_{34}^2}{x_{13}^2 x_{14}^2}\right)\right) I_{x_1 x_3}^2 I_{x_1 x_4} \delta_{\mu ,\nu}\nonumber\\
&+& 2 \pi ^2 x_{13,\mu } x_{34,\nu } I_{x_1 x_3}^2 I_{x_1 x_4} I_{x_3 x_4} +2 \pi ^2  x_{14,\mu } x_{34,\nu }I_{x_1 x_3} I_{x_1 x_4}^2 I_{x_3 x_4}\,, \nonumber 
\>
\<
\lim_{x_2\rightarrow x_1}\partial_{3,\mu}\partial_{4,\nu}X_{1234}&\;=\;& \frac{\delta_{\mu \nu}}{2} I_{x_1x_3}I_{x_1x_4}I_{x_3x_4} - 4 \pi^2 x_{34, \mu} x_{34, \nu} I_{x_1x_3}I_{x_1x_4}I^2_{x_3x_4} \nonumber \\ 
&-& 4 \pi^2 x_{14, \nu} \, x_{34, \mu} \, I_{x_1x_3}I^2_{x_1 x_4} I_{x_3 x_4} + 4 \pi^2 x_{13, \mu} \, x_{34, \nu} \, I^2_{x_1 x_3}I_{x_1 x_4} I_{x_3 x_4}  \nonumber \\
&-& 4 \pi^2 x_{13,\mu} \, x_{14,\nu} \, \log\left( \frac{\epsilon^2 x_{34}^2}{x_{13}^2 x_{14}^2}\right) I^2_{x_1x_3}I^2_{x_1x_4} \, .\nonumber
\>

\indent Using the above results, we can proceed to the computation of the two- and three-point functions. The result of all the non-zero Feynman diagrams relevant for us is given in figure \ref{fulldetails}, where we have omitted terms involving  $\epsilon^{\mu \nu \rho \lambda}$ that must either vanish when a pair of point collide or cancel when all the diagrams are summed. This is the case in order to preserve conformal invariance and parity.

The results of figure \ref{fulldetails} only contain derivatives of the function $H_{12,34}$ and it is possible to evaluate them explicitly \cite{Georgiou:2004ty}.  Consider the case when the derivatives act on either the first or the second pair of points of $H$, namely $\partial_1\cdot \partial_2 H_{12,34}$, and also the case when they act on a point belonging to the first pair and a point belonging to the second pair, for instance $\partial_1\cdot \partial_4 H_{12,34}$. The first case is straightforward to compute by using integration by parts and the property of the euclidean propagator $\Box_{x}I_{xy}=-\delta^{(4)}(x-y)$. The result is 
\[
\partial_{1}\cdot\partial_{2}H_{12,34}=\frac{1}{2}\left(Y_{134}I_{x_1 x_2}+Y_{234}I_{x_1 x_2}-X_{1234}\right)\label{d12H}\,.
\]
For computing the second case, we need the function $F_{12,34}$ defined in (\ref{F}) and some identities of $H_{12,34}$. Firstly, note that $H$ satisfies the equation
\[
\left(\partial_{1,\mu}+\partial_{2,\mu}+\partial_{3,\mu}+\partial_{4,\mu}\right)H_{12,34}=0 \, , \label{trans}
\]
which can be proved by integration by parts. Similarly, it is possible to show that the following identity holds
\[
\partial_i \cdot \partial_j H_{12,34}=\frac{1}{2}\left(\Box_k+\Box_l -\Box_i -\Box_j\right) H_{12,34}+\partial_k \cdot \partial_l H_{12,34}\label{box}
\] for $i\neq j\neq k\neq l$. In order to get $\partial_1\cdot \partial_4 H_{12,34}$, it is convenient to write it as
\[
\partial_1 \cdot \partial_4 H_{12,34}=\frac{1}{2}\left(\partial_1\cdot \partial_3 +\partial_1 \cdot \partial_4 \right)H_{12,34}-\frac{1}{2}\left(\partial_1\cdot \partial_3 -\partial_1 \cdot \partial_4 \right)H_{12,34} \,. \label{ddH}
\]Now using (\ref{box}), one can show that the first term on the right-hand side of (\ref{ddH}) can be written as

\begin{figure}[t!]
\begin{center}
\includegraphics[width=1\linewidth]{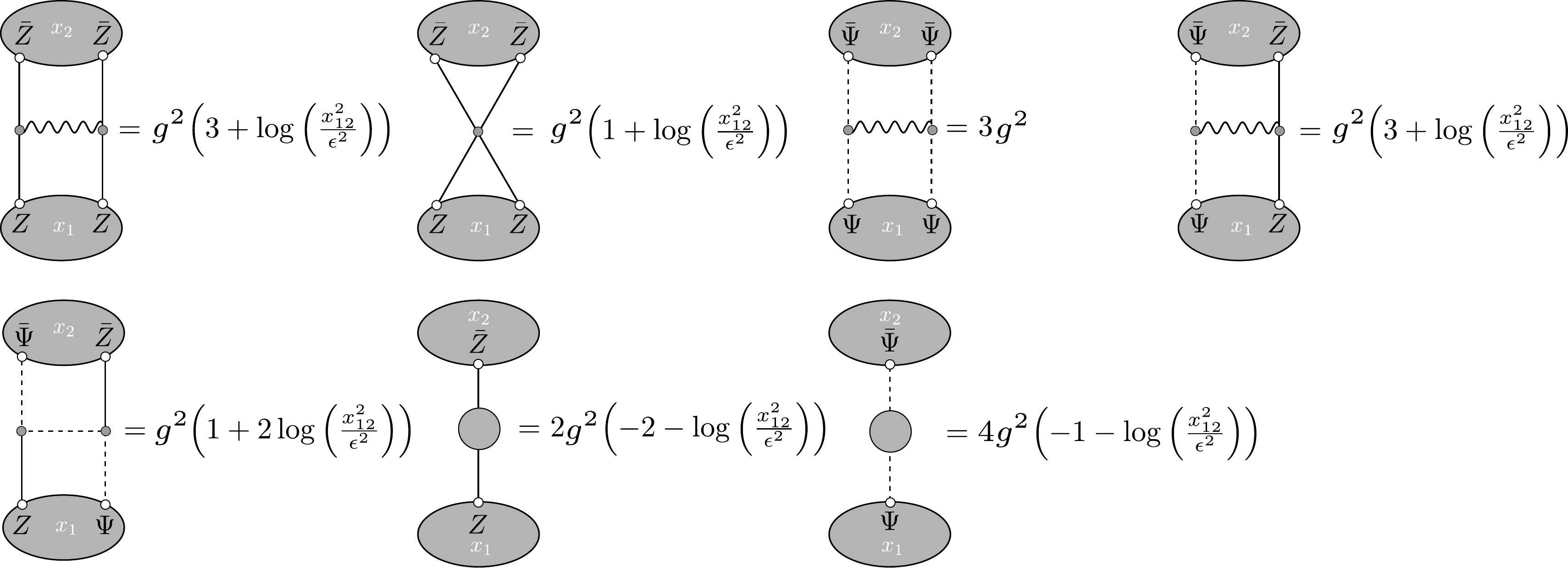}
\caption{The results of the Feynman diagrams for the two-point functions. They are obtained by taking the limits $x_3\rightarrow x_2$ and $x_4\rightarrow x_1$ of the expressions in figure \ref{fulldetails}. \label{2pt} }
\end{center}
\end{figure}
\begin{figure}[t!]
\begin{center}
\includegraphics[width=1.0\linewidth]{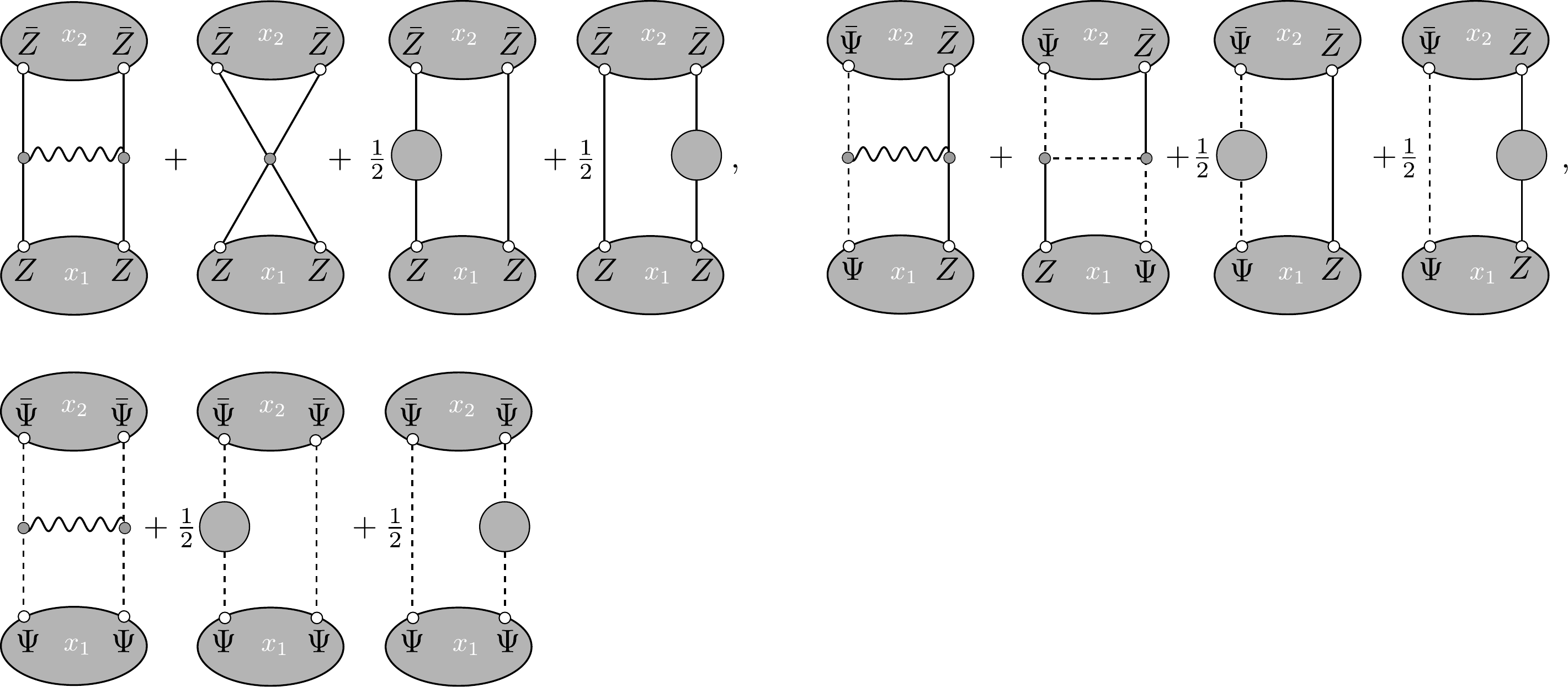}
\caption{Using the results of figure \ref{2pt}, the sum of the graphs appearing in this figure gives precisely the $\alg{su}(1|1)$ Hamiltonian of
(\ref{1loopH}).}\label{su11ham}
\end{center}
\end{figure}
\[
\frac{1}{2}\left(\partial_1\cdot \partial_3 +\partial_1 \cdot \partial_4 \right)H_{12,34}=-\frac{1}{2}\left(\Box_1 H_{12,34}+\partial_1 \cdot \partial_2 H_{12,34}\right) \label{inter1}
\]
where $\Box_i H_{12,34}$ can be computed using the equation defining the euclidean propagator and $\partial_1 \cdot \partial_2 H_{12,34}$ is known from (\ref{d12H}). Using (\ref{trans}), the second term on the right-hand side of (\ref{ddH}) can be written as
\[
\frac{1}{2}\left(\partial_1\cdot \partial_3 -\partial_1 \cdot \partial_4 \right)H_{12,34}=\frac{1}{4}\left(F_{12,34}I_{x_1 x_2} I_{x_3 x_4}+(\Box_4 -\Box_3)H_{12,34}\right)\,. \label{inter2}
\]
Finally, substituting (\ref{inter1}) and (\ref{inter2}) in (\ref{ddH}), one gets an expression for $\partial_1\cdot \partial_4 H_{12,34}$.  The expressions for the remaining cases where the derivatives act on other points  can be deduced analogously.\\
\indent In order to get the two-point functions, one takes the limit where two pairs of points collapse into each other that is  $x_4\rightarrow x_1$ and $x_{3}\rightarrow x_{2}$. The results of these limits are given 
in figure 
\ref{2pt}. Summing all the diagrams as illustrated in figure 
\ref{su11ham}, one obtains the one-loop 
Hamiltonian operator
\[
H=2g^2 (I-SP) \label{1loopH}\,,
\] 
where $SP$ is the superpermutator which exchanges the fields and picks up a minus sign when both fields are fermionic. This Hamiltonian is the well known result of \cite{BeisertTwoloop,StaudacherTwoloop}. \\
\indent We now proceed to the three-point functions. In order to obtain the constant coming from each diagram, one takes the limit  of the expressions  given in figure \ref{fulldetails} where a single pair of points collapses into each other. After taking that limit, the result  will have constant terms, divergent logarithmic terms and eventually $Y$ functions and their derivatives. The derivatives of the $Y$ functions can be expressed in terms of the $Y$ function itself by using some of its properties. This will be explained in detail in the Appendix \ref{examples}. After this procedure, the logarithmic terms will contribute to the standard regulator dependence in (\ref{eq:3ptbare}) and the remaining $Y$ functions will cancel with similar contributions from other diagrams in a way that the conformal invariance is restored. One can then read the constant part of the diagram. The final step is to subtract one half the constant coming from the same diagram but when the two pairs of points collapse into each other as described in figure \ref{3ptdiagram}. The results are given in figure \ref{results}. 

\begin{figure}[t!]
\begin{center}
\includegraphics[width=0.7\linewidth]{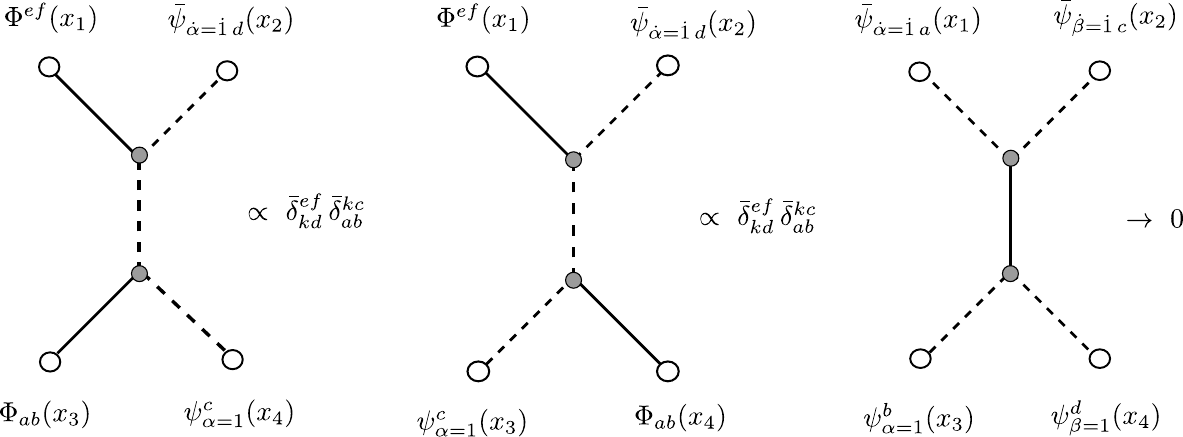}
\caption{The additional Feynman diagrams that do not contribute in the setup considered in this work. The solid and dashed lines represent the scalars and fermions, respectively.\label{newgraphic}}
\end{center}
\end{figure}

Let us comment now on a detail of this computation. Our final results presented in the figures \ref{results} and \ref{fulldetails} do not contain the Feynman diagrams of figure \ref{newgraphic}. This is the case because the first 
two graphs of this figure turn out to cancel among them. 
They can give a non-zero contribution in our setup only when either $b=c=4$ and $a=3$ or $a=c=2$ and $b=3$. However, as these two graphs always appear with the same weight and opposite signs, they end up canceling. The last graph
of the figure \ref{newgraphic} must vanish 
when $|x_{12}| \rightarrow 0$ because
it is 
\begin{equation}
\propto \, \, \epsilon^{\gamma \delta} \epsilon^{\dot{\gamma} \dot{\delta}} \sigma^{\mu}_{E \gamma \dot{1}} \sigma^{\nu}_{E \delta \dot{1}}
\sigma^{\rho}_{E 1 \dot{\gamma}} \sigma^{\lambda}_{E 1 \dot{\delta}} \int d^4 v \, d^4 u \, (\partial^{v}_{\mu}I_{x_1 v})(\partial^v_{\nu}I_{x_2 v})
I_{vu}
(\partial^{u}_{\rho}I_{x_3 u})(\partial^u_{\lambda}I_{x_4 u}) \, \rightarrow 0 \, .  \nonumber 
\end{equation}  
If it was non-vanishing it would
produce a term with a different tensor structure of (\ref{eq:3ptbare}) which would violate conformal invariance.

\section{Some examples of three-point functions}\label{examples}
\begin{figure}[t!]
\begin{center}
\includegraphics[width=0.35\linewidth]{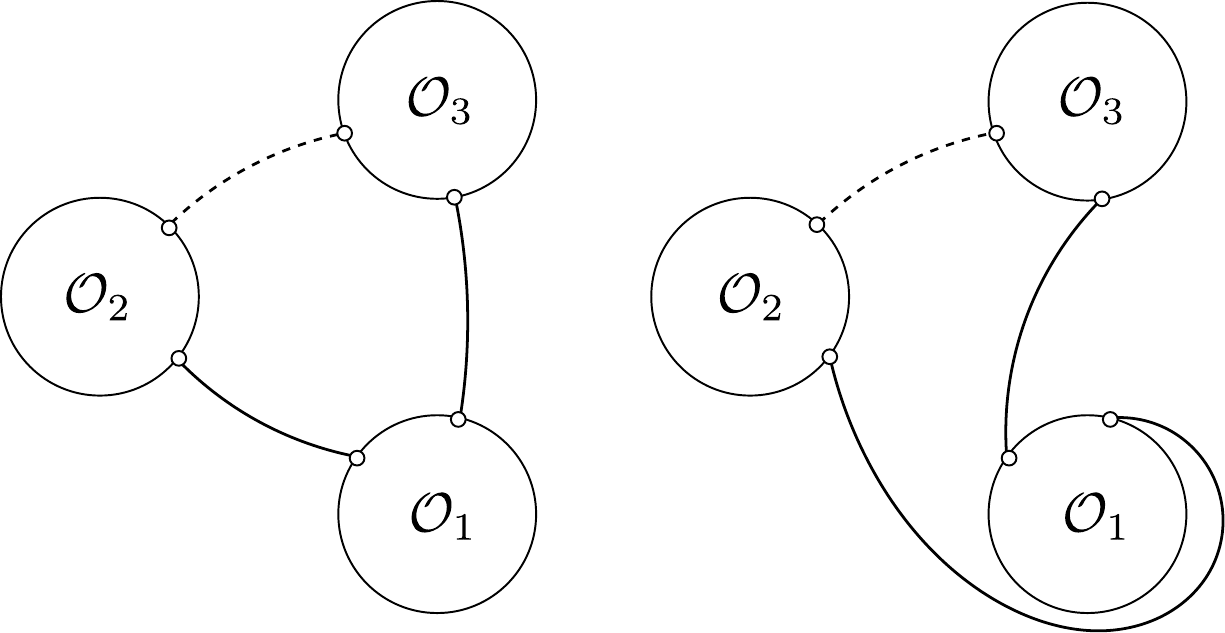}
\caption{The tree-level diagrams for the three-point functions of the three half-BPS operators considered in (\ref{O1bps})-(\ref{O3bps}). Note that only the first term of $\mathcal{O}_3$ in (\ref{O3bps}) gives a non-zero contribution at this order in perturbation theory as the second term clearly gives a vanishing contribution due to $R$-charge conservation. }\label{tree}
\end{center}
\end{figure}
In this Appendix, we give two examples of three-point functions. The first one is the case of three half-BPS operators. It is well-known that this correlator is protected and therefore it constitutes a check for our computations. Then we compute a non-protected three-point function both by brute force and by using our prescription of inserting the operator $\mathcal{F}$ at the splitting points.

\subsection{Three half-BPS operators} \label{3BPS}
\begin{figure}[t!]
\begin{center}
\includegraphics[width=1.0\linewidth]{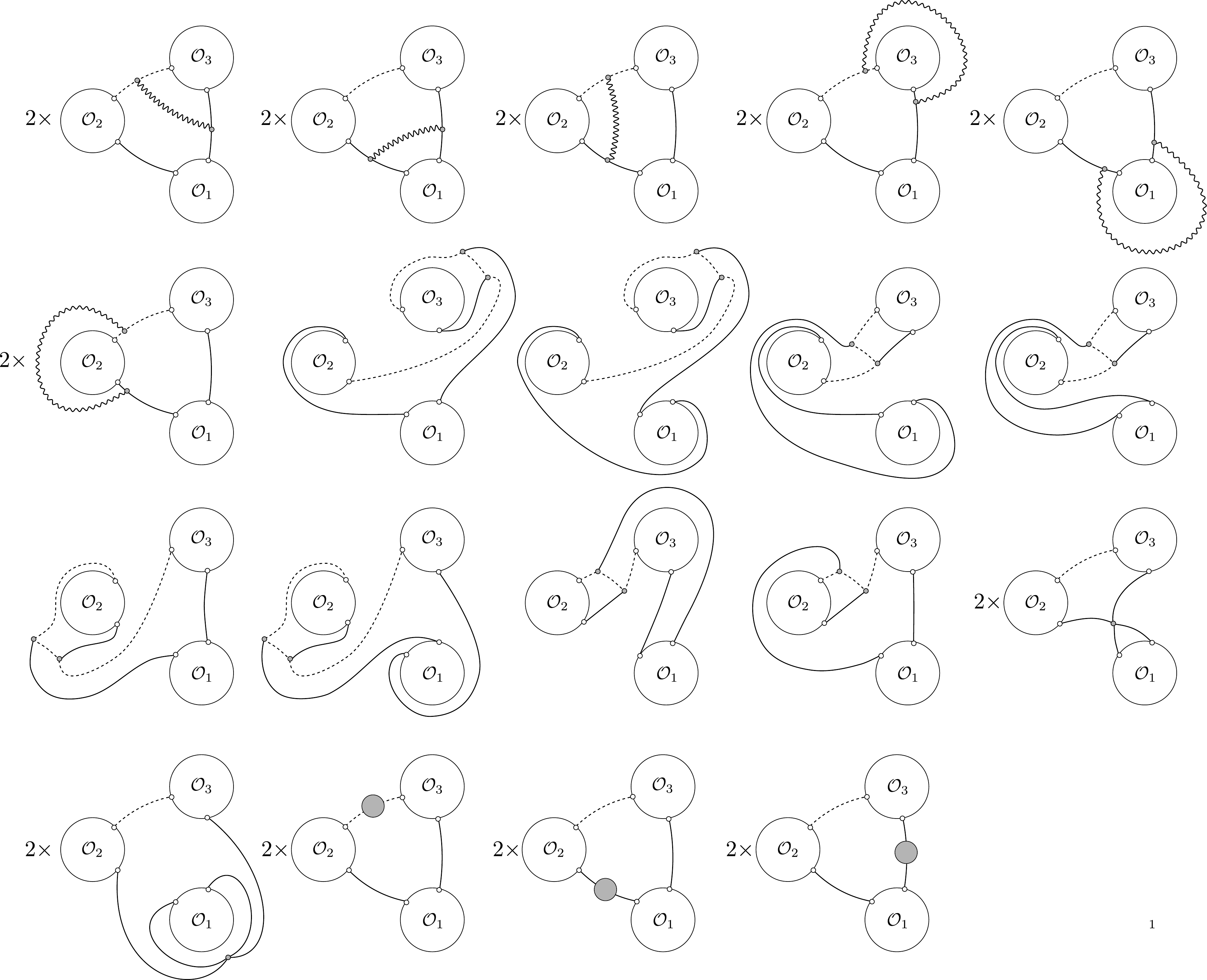}
\caption{The one-loop diagrams contributing to the three-point function of the three half-BPS operators considered in (\ref{O1bps})-(\ref{O3bps}). In the last four diagrams of the second row, the second term of $\mathcal{O}_3$ (see expression (\ref{O3bps})) gives a non-zero result. In all other diagrams only the first term of $\mathcal{O}_3$ contributes.}\label{1loop}
\end{center}
\end{figure}

Consider the following three half-BPS operators
\<
\mathcal{O}_1\;&=&\;\text{Tr}\left( Z Z \right) \, , \label{O1bps}\\
\mathcal{O}_2\;&=&\;\text{Tr} \left( \bar{\Psi}\bar{Z} \right) \, ,\\
\mathcal{O}_3\;&=&\;(\alg{R}^{2}_{\;\;4 }\alg{R}^{1}_{\;\;3 }) \cdot\text{Tr}\left(  \Psi  Z\right)=\text{Tr}\left(\Psi \bar{Z}\right)+\text{Tr} \left( \psi^{2}
\,\Phi^{14} \right)  \label{O3bps}\,.
\>
At tree-level the result is simply given by the sum of the two diagrams of figure \ref{tree} and reads
\[
\langle \mathcal{O}_1 (x_1) \mathcal{O}_2 (x_2) \mathcal{O}_3 (x_3) \rangle = \frac{2}{ (2 \pi)^6 8^2 x_{12}^2 x_{13}^2} (\sigma^{\mu}_{E})_{1 \dot{1}}\partial_{3,\mu}\frac{1}{2 x_{23}^2} \, . 
\]
At one-loop, one has to sum the diagrams of figure \ref{1loop} and use the results given in figure \ref{fulldetails}  taking the appropriate limits.
\begin{figure}[t!]
\begin{center}
\includegraphics[width=0.9\linewidth]{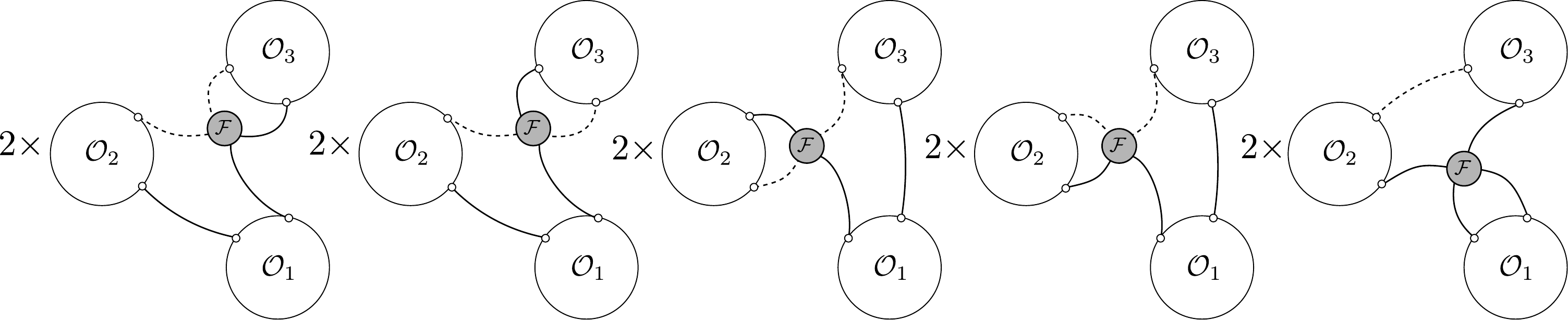}
\caption{Inserting the $\mathcal{F}$ operator at the splitting points, one reproduces the vanishing result expected for a three-point function of half-BPS operators. Apart from the graphics in the figure, there are similar graphics with the operator $\mathcal{F}$ acting in the remaining splitting points.}\label{insertions}
\end{center}
\end{figure}Some diagrams will still contain the function $Y$ and its first derivatives. The $Y$ function depends on the external points in a way that does not respect the spacetime dependence fixed by conformal symmetry, see equation (\ref{eq:3ptbare}). However, when one sums  the different diagrams this non-conformal spacetime dependence turns out to cancel identically.\\
\indent At the end of the day, summing all the diagrams gives a vanishing one-loop contribution to the structure constant in agreement with the non-renormalization theorem for the three-points functions of half-BPS operators introduced in \cite{Minwalla}.
Equivalently, one can also use our prescription to reproduce this one-loop result. One simply has to sum over the insertions represented in figure \ref{insertions} obtaining zero as expected.
\begin{figure}[t]
\begin{center}
\includegraphics[width=0.35\linewidth]{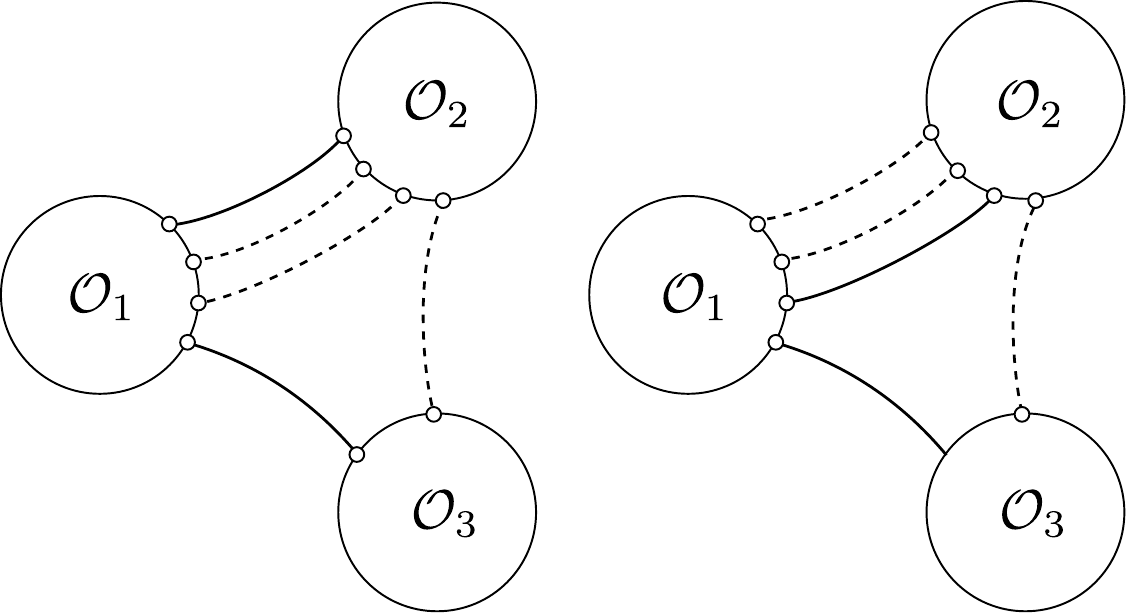}
\caption{The tree-level diagrams contributing to the three-point function of the operators (\ref{nonBPS1})-(\ref{O3nonbps}). Once again, we only need to consider the first term of (\ref{O3nonbps}) at this order in perturbation theory.}\label{treenon}
\end{center}
\end{figure}
\subsection{Two non-BPS and one half-BPS operators}
\begin{figure}[H]
\begin{center}
\includegraphics[width=1\linewidth]{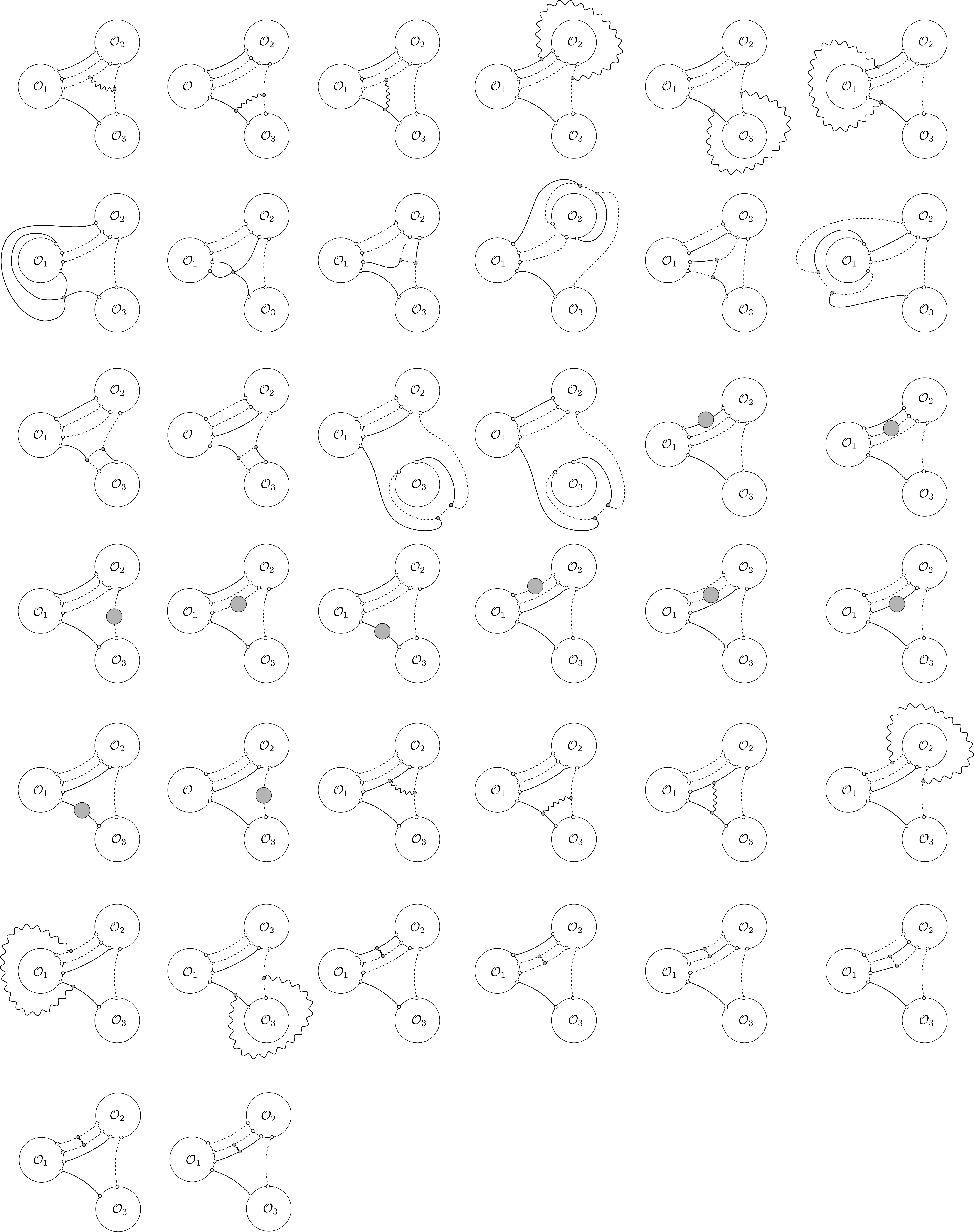}
\caption{The relevant one-loop diagrams for the three-point function of the operators  (\ref{nonBPS1})-(\ref{O3nonbps}). The second term of (\ref{O3nonbps}) gives a non-zero contribution namely the first four diagrams of the third row.}\label{onenon}
\end{center}
\end{figure}

We consider now a non-protected three-point function. This example serves as an illustration of some of the technical details of the brute force computation. Moreover, we also use it to check our prescription of the $\mathcal{F}$ operator insertion at the splitting points. The operators at one-loop level that we will consider are 
\<
\mathcal{O}_1\;&=&\;\text{Tr}\left(Z \Psi \Psi Z\right) \, , \label{nonBPS1}\\
\mathcal{O}_2\;&=&\;\text{Tr}\left( \bar Z \bar\Psi \bar\Psi \bar\Psi  \right) \, , \\
\mathcal{O}_3\;&=&\;(\alg{R}^{2}_{\;\;4 }\alg{R}^{1}_{\;\;3 }) \cdot\text{Tr}\left(  \Psi  Z\right)=\text{Tr}\left(\Psi \bar{Z}\right)+\text{Tr} \left( \psi^{2}
\,\Phi^{14} \right) \label{O3nonbps}\,.
\>
Note that the $\mathcal{O}_1$ and $\mathcal{O}_2$ are not half-BPS and therefore they will receive corrections as explained in section \ref{two loop}. However, to compute the Feynman diagrams contribution we do not need to take them into account.
At tree-level the result is simply the sum of the two diagrams of figure 
\ref{treenon} which gives
\<
\langle \mathcal{O}_1 (x_1) \mathcal{O}_2 (x_2) \mathcal{O}_3 (x_3) \rangle_0 =  - \frac{2}{(2 \pi)^{10} 8^2 x_{13}^2 
x_{12}^2} \nonumber \hspace{60mm} \\ \times \left( (\sigma^{\mu}_E)_{1 \dot{1}}\partial_{1,\mu}\frac{1}{2x_{12}^2} \right) \left( (\sigma^{\nu}_E)_{ 1 \dot{1}}\partial_{1,\nu} \frac{1}{2x_{12}^2} \right) \left( (\sigma^{\rho}_E)_{1 \dot{1}}\partial_{3,\rho} \frac{1}{
2x_{23}^2} \right)\,\nonumber.
\>
The diagrams contributing at one-loop are represented in figure 
\ref{onenon}.

As in the previous example, the dependence of each diagram on the $Y$ function and its derivatives will cancel when we sum over all the diagrams. This ensures that we obtain a conformal invariant result. However, this cancellation is not immediate and it relies on several properties of the $Y$ function.
The first observation is that the function $Y$ is given by 
\[
Y_{123}=\frac{\pi^2 \phi(r,s)}{(2\pi)^4}I_{x_1,x_3}, \label{YandPhi}
\]
\begin{figure}[H]
\begin{center}
\includegraphics[width=0.95\linewidth]{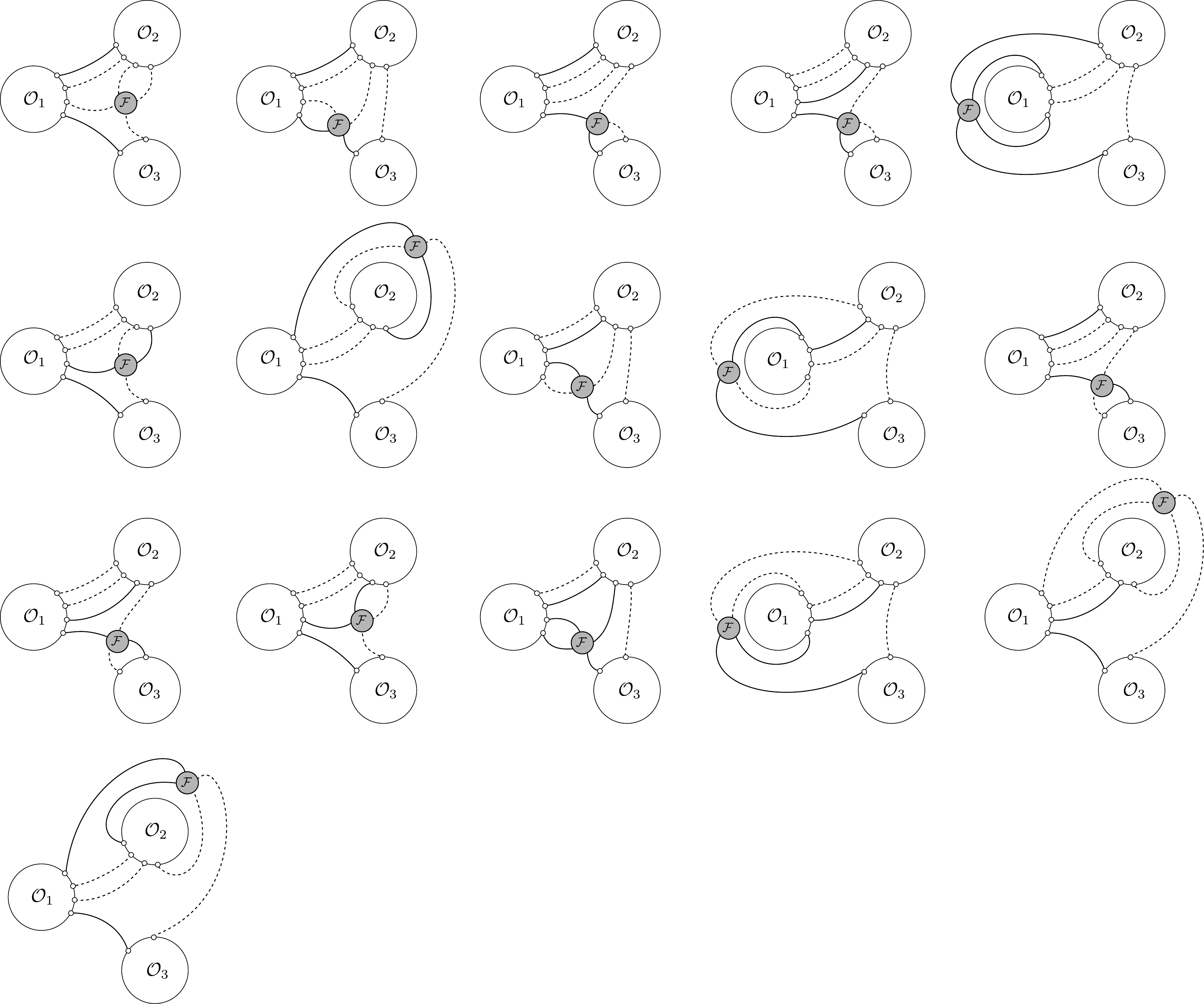}
\caption{Inserting the operator $\mathcal{F}$ at the splitting points reproduce the result of the one-loop Feynman diagrams.}\label{insertions2}
\end{center}
\end{figure} 
\noindent where $r=\frac{x_{12}^2}{x_{13}^2}$ and $s=\frac{x_{23}^2}{x_{13}^2}$ and an explicit expression for $\phi(r,s)$ can be found in \cite{BeisertBMN}. The important information for us is that the function $\phi$ satisfies the following differential equations \cite{Alday}
\<
&&\phi(r,s)+(s+r-1)\partial_s \phi(r,s)+2 r \partial_r\phi(r,s)=-\frac{\log r}{s} \, ,\label{Phiidentities}\\
&&\phi(r,s)+(s+r-1)\partial_r \phi(r,s)+2 s \partial_s\phi(r,s)=-\frac{\log s}{r} \, , \nonumber
\>
which can be used to relate the first derivatives of $Y$ with $Y$ itself. In addition, one can take derivatives with respect to $r$ and $s$ of both the equations above to arrive at a system of equations that relates second derivatives of $\phi$ with first derivatives and the function $\phi$ itself. Using then (\ref{YandPhi}), it is trivial to get rid of the second derivatives of $Y$. These properties of
the function $\phi$ ensure that the non-conformal dependence of the three-point function indeed cancel when all diagrams are summed over. 

The final result  is given by
\[
\langle \mathcal{O}_1(x_1) \mathcal{O}_2(x_2) \mathcal{O}_3(x_3) \rangle=\langle \mathcal{O}_1  \mathcal{O}_2  \mathcal{O}_3  \rangle_0 \left(1+4 g^2\left(-1+2\log \left( \frac{\epsilon^2}{x_{12}^2}\right)\right)+\mathcal{O}(g^4) \right) \nonumber
\]
which comparing with (\ref{eq:3ptbare}) gives the correct anomalous dimensions of the operators. This is a non-trivial consistency check of our computation.\\
\indent  The structure constant can now be obtained by also computing the constants from the two-point functions. We have all the tools at hand to perform such calculation and read the one-loop constant. We obtain the following contribution from the Feynman diagrams to the structure constant
\[
C_{123}^{(1)}\bigg|_{\text{Feynman diagrams contribution}}=-\frac{3}{2} \, . \label{resultC}
\]
Recall that this is not the final result, one also has to add the extra contribution from the corrected two-loop Bethe states.\\
\indent Finally, it is possible to test our prescription of inserting the $\mathcal{F}$ operator at the splitting points, see figure 
\ref{insertions2}. Summing over all these insertions gives precisely the contribution (\ref{resultC}) to the structure constant.

\section{Wilson line contribution} \label{Wline}
\begin{figure}[t!]
\begin{center}
\includegraphics[width=0.98\linewidth]{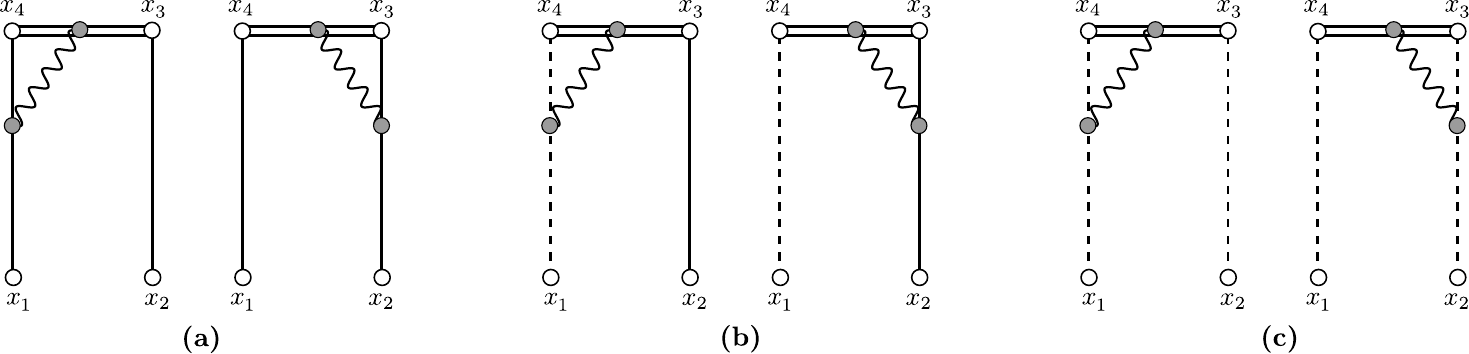}
\caption{The one-loop additional graphs coming from the Wilson line.}\label{WL}
\end{center}
\end{figure}
\begin{figure}[t!]
\begin{center}
\includegraphics[width=0.6\linewidth]{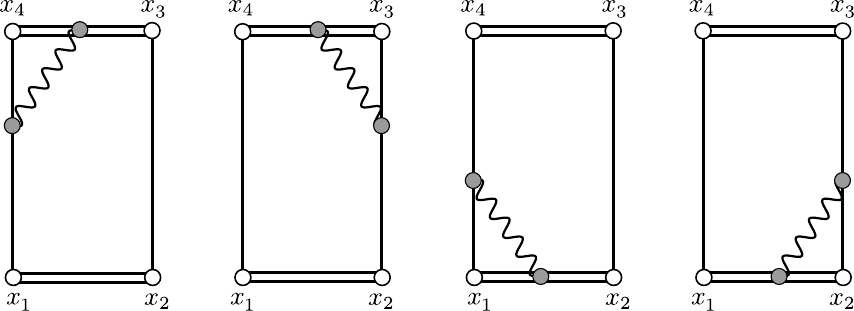}
\caption{The Wilson line contributions to the two-point functions. In the combination of  the two- and the three-point diagrams that provide the scheme independent structure constant (\ref{eq:3ptbare}), all the extra diagrams coming from the Wilson lines cancel each other at this order in perturbation theory. In the figure, the diagram corresponding to the emission of a gluon between the two Wilson lines is not depicted, since it is proportional to $\epsilon^2$ and vanishes in the limit $\epsilon \rightarrow 0$.   }\label{WL2pt}
\end{center}
\end{figure}
As mentioned before, the point splitting regularization breaks explicitly the gauge invariance due to the fact that some fields inside the trace are now at a slightly different spacetime points. The introduction of a Wilson line connecting these fields restore the gauge invariance at the price of introducing extra Feynman diagrams. In this Appendix, we show that these extra diagrams do not contribute to the scheme and normalization independent structure constant $C^{(1)}_{123}$ defined in (\ref{eq:3ptbare}).

\subsection{Wilson line connecting two scalars}

In our conventions the Wilson line
operator is defined by
 \begin{equation}
W_l = \mathcal{P} \, \text{exp}\left[ \,  i \, g_{YM} \int A_{\mu} \, d \vec{x}^{\, \mu} \right] \, . \nonumber
\end{equation}
When inserting a Wilson line connecting two scalars, it is necessary to consider the one-loop graphs corresponding to the gluon emission depicted in figure \ref{WL}(a). 
Let us define $\epsilon^{\mu} = x_4^{\mu}-x_3^{\mu}$ and at the end of the day we will take the limit $\epsilon^{\mu} \rightarrow 0$. Then we can conveniently parametrize the Wilson line by $x^{\mu}(z)=x_3^{\mu}+z \epsilon^{\mu}$. The result of the sum of the diagrams is 
\[
\text{figure\,} \ref{WL}(a) = \frac{\lambda}{128} \int_{0}^{1} dz\, \epsilon_{\mu} \left( I_{x_2 x_3} \, \partial_{1}^{\mu}\, Y_{1 x 4} - I_{x_1 x_4} \,\partial_{2}^{\mu}\, Y_{2 x 3} +
I_{x_1 x_4} \, \partial_{3}^{\mu}\, Y_{2 x 3} -
I_{x_2 x_3} \, \partial_{4}^{\mu}\, Y_{1 x 4} \right) \, , \label{WLescalar}
\]
where we have suppressed both the $R$-charge and the gauge indices which are the same as in the tree-level case. From the formula (\ref{dY}), it follows that the first and second terms of the above result are of order $\epsilon$ and therefore vanish in the limit $\epsilon\rightarrow 0$. However, from (\ref{oneovereps}) we see that the third and last term give a finite contribution. 

In order to compute the scheme and normalization structure constant $C^{(1)}_{123}$ of (\ref{eq:3ptbare}),
we have to subtract from the previous result one half of the one-loop diagrams from the two-point functions as shown in figure 
\ref{WL2pt} (we take both the limits $x_4 \rightarrow x_3$ and $x_2 \rightarrow x_1$). 
It is simple to show that the contribution of these diagrams cancels exactly the constant coming from the expression (\ref{WLescalar}). So, at this order in perturbation theory we do not get any further contribution to $C^{(1)}_{123}$ and therefore we can safely ignore the Wilson lines.

\subsection{Wilson line connecting either a scalar and a fermion or two fermions}
In the case of a scalar and a fermion connected by a Wilson line, the contribution of the diagrams depicted in \ref{WL}(b) is given by
\<
\text{figure\,} \ref{WL}(b) = \frac{\lambda}{32} \int_{0}^{1} dz\, \epsilon^{\mu}  \Bigl[ I_{x_2x_3}( \sigma_{E \mu 1 \dot{1}} \partial^4 \cdot \partial^1 Y_{1x4} - \sigma^{\nu}_{E 1 \dot{1}} \partial^4_{\nu}\partial^1_{\mu}Y_{1x4}-\sigma^{\nu}_{E 1 \dot{1}} \partial^4_{\mu} \partial^1_{\nu}Y_{1x4}  \nonumber \\
- \epsilon_{\rho \mu \lambda \nu} \, \sigma^{\nu}_{E 1 \dot{1}} \, \partial^{4,\lambda} \partial^{1,\rho} Y_{1x4})+ \sigma^{\nu}_{E 1 \dot{1}} \partial^{1}_{\nu}I_{x_1 x_4} (\partial^3_{\mu}Y_{2x3}-\partial^{2}_{\mu}Y_{2x3})  \,\Bigr]\, . \hspace{15mm} \label{WLscalarfermion}
\>
Using the expressions (\ref{oneovereps}-\ref{lastY}), one can easily see that this gives a finite contribution in the limit when $\epsilon$ goes to zero (in particular, the term with $\epsilon_{\rho \mu \lambda \nu}$ vanishes). To this result, we have again to subtract one half of the one-loop diagrams from the two-point functions as was done in the previous subsection for the case of two
scalars. Once again, the contribution of these diagrams cancels exactly the expression (\ref{WLscalarfermion}). 

In the case when we have a Wilson line connecting two fermions, the same argument holds. Hence at one-loop level, we can ignore the Wilson lines contributions in all cases.

\section{A note on the $\alg{su}(1|1)$ invariance of the final result} \label{invariance}
In this Appendix, we address the question of the $\alg{su}(1|1)$ invariance of our formula for the structure constant. In particular, this serves as a consistency check for the one-loop prescription we have computed.\\
\indent Let us start by checking the tree-level structure constant. Its expression is given in (\ref{final}) when $g\rightarrow0$.
One possible way of implementing a symmetry transformation on a state at any value of the coupling is to add a Bethe root with 
zero-momentum. It is clear from the Bethe equations that we obtain a state with the same energy and therefore belonging to the same multiplet as the original one. Consider then the states $|\bold{1}\rangle$ and $|\bold{2}\rangle$  with one of their
momenta $p^{(1)}_j$ and $p^{(2)}_i$ being equal to zero.
In this particular case, we can write    
apart from possible signs

\begin{equation}
| \bold{2} \rangle =  \, \, \bar{\alg{Q}}  \, | \bold{2}, \{ \hat{p}^{(2)}_i \} \rangle \, , \quad \quad \langle \bold{1}^f | =  \, \, \langle \bold{1}^f, \{ \hat{p}^{(1)}_j \} | \, \bar{\alg{S}} \,, \nonumber
\end{equation}
where the hat over a $p$ means that this momentum is absent. The operator $\bar{\alg{Q}}$ ($\bar{\alg{S}}$)  creates (annihilates) a zero-momentum magnon on a ket and annihilates (creates) a zero-momentum magnon on the bra (we are omitting the $R$-charge and Lorentz indices for simplicity).  

For this particular choice of momenta the expression (\ref{final}) for $g=0$ becomes
\<
\langle \bold{1}^f, \{ \hat{p}^{(1)}_j \} | \, \bar{\alg{S}} \, \,  \tilde{\mathcal{O}}_3 \, \, \bar{\alg{Q}}  \, | \bold{2}, \{ \hat{p}^{(2)}_i \} \rangle = \langle \bold{1}^f, \{ \hat{p}^{(1)}_j \} |  \,  \tilde{\mathcal{O}}_3 \, \, \{ \bar{\alg{S}} , \bar{\alg{Q}} \}  \, | \bold{2}, \{ \hat{p}^{(2)}_i \} \rangle \, , \label{treeaindaone}
\>
where we denote the operator $|\bar{Z}\dots\bar{Z} i_1 \ldots i_{L_2-N_3} \rangle \langle \bar\Psi\dots \bar\Psi i_1 \ldots i_{L_2-N_3}  |$ by $\tilde{\mathcal{O}}_3$. Moreover in this equality, we have 
used that $\bar{\alg{S}}$ and $\tilde{\mathcal{O}}_3 $ commute
which can be proved by applying the commutator to a generic $\alg{su}(1|1)$ state. In addition, we have also used that  
\begin{equation}
\bar{\alg{S}} \, \, | \bold{
2}, \{ \hat{p}^{(2)}_i \} \rangle = 0 \, ,
\end{equation}
as the state is primary. 

Now, the anticommutator $ \{ \bar{\alg{S}} , \bar{\alg{Q}} \}$  is given by (see for instance the Appendix D of \cite{BeisertThesis})
\[
 \{ \bar{\alg{S}} , \bar{\alg{Q}} \} = \mathcal{L}+\frac{1}{2}H(g) \, ,
\]
where $\mathcal{L}$ is the length operator and $H(g)$ is the dilatation operator. When acting on the state 
$| \bold{2},\{ \hat{p}^{(2)}_i \} \rangle$ it gives at leading order the length of the state $L_2$. In conclusion, we have derived the following equality
\begin{equation}
| \, \langle \bold{1}^f  | \, 
\tilde{\mathcal{O}}_3 \, |  \bold{2} \rangle \, | = L_2 \, | \, \langle \bold{1}^f, \{ \hat{p}^{(1)}_j \} |  \,  \tilde{\mathcal{O}}_3 \,  | \bold{2}, \{ \hat{p}^{(2)}_i \} \rangle \,  | \,. \label{equalitytree}    
\end{equation}  

The relation above shows how the structure constant changes under $\alg{su}(1|1)$ transformations of the states $| \bold{2} \rangle$ and $ \langle \bold{1}^f |$ at leading order. It is now simple to check that our expression 
for this scalar product given in the main text indeed satisfies this relation.

At one-loop, the final expression for the structure constants given in (\ref{final}) has the following 
term
\<
\langle \bold{1}^f | \, \tilde{\mathcal{O}}'_3 \,| \bold{2} \rangle \, , 
\>where
\begin{equation}
\tilde{\mathcal{O}}'_3 = \left(1+ \frac{g^2}{2}\mathcal{F}_{L_3-N_3,L_3-N_3+1}  +\frac{g^2}{2} \mathcal{F}_{L_{1}, 1} \right) \,  \tilde{\mathcal{O}}_3  
\, \left(1+ \frac{g^2}{2} \mathcal{F}_{N_{3},N_{3}+1} +\frac{g^2}{2}\mathcal{F}_{L_{2}, 1}\right) \, .
\end{equation}

If we require that $\tilde{\mathcal{O}}'_3$ 
commutes with the generator $\bar{\alg{S}}$ then we find that at one-loop the relation (\ref{equalitytree}) becomes   
\begin{equation}
| \, \langle \bold{1}^f  | \, 
\tilde{\mathcal{O}}'_3 \, |  \bold{2} \rangle \, | = \left(L_2 + \frac{1}{2} \gamma_2 \right) \, | \, \langle \bold{1}^f, \{ \hat{p}^{(1)}_j \} |  \,  \tilde{\mathcal{O}}'_3 \,  | \bold{2}, \{ \hat{p}^{(2)}_i \} \rangle \,  | \,,    
\end{equation}  
where $\gamma_2$ is the one-loop anomalous dimension of the operator $\mathcal{O}_2$. We have verified 
that 
this relation is indeed obeyed, which shows that our prescription respects these symmetry constraints.


\begin{thebibliography}{1}




\bibitem{Spectrum} 
  N.~Gromov, V.~Kazakov and P.~Vieira,
  ``Exact Spectrum of Anomalous Dimensions of Planar N=4 Supersymmetric Yang-Mills Theory,''
  Phys.\ Rev.\ Lett.\  {\bf 103}, 131601 (2009)
  [arXiv:0901.3753 [hep-th]].



\bibitem{BigReview} 
  N.~Beisert, C.~Ahn, L.~F.~Alday, Z.~Bajnok, J.~M.~Drummond, L.~Freyhult, N.~Gromov and R.~A.~Janik {\it et al.},
  ``Review of AdS/CFT Integrability: An Overview,''
  Lett.\ Math.\ Phys.\  {\bf 99}, 3 (2012)
  [arXiv:1012.3982 [hep-th]].


\bibitem{pmu} 
  N.~Gromov, V.~Kazakov, S.~Leurent and D.~Volin,
  ``Quantum spectral curve for $AdS_5/CFT_4$,''
  Phys.\ Rev.\ Lett.\  {\bf 112}, 011602 (2014)
  [arXiv:1305.1939 [hep-th]].



\bibitem{TailoringI} 
  J.~Escobedo, N.~Gromov, A.~Sever and P.~Vieira,
  ``Tailoring Three-Point Functions and Integrability,''
  JHEP {\bf 1109}, 028 (2011)
  [arXiv:1012.2475 [hep-th]].



\bibitem{FodaMetodo} 
  O.~Foda,
  ``N=4 SYM structure constants as determinants,''
  JHEP {\bf 1203}, 096 (2012)
  [arXiv:1111.4663 [math-ph]].




\bibitem{Japas} 
  K.~Okuyama and L.~-S.~Tseng,
  ``Three-point functions in N = 4 SYM theory at one-loop,''
  JHEP {\bf 0408}, 055 (2004)
  [hep-th/0404190].



\bibitem{Alday} 
  L.~F.~Alday, J.~R.~David, E.~Gava and K.~S.~Narain,
  ``Structure constants of planar N = 4 Yang Mills at one loop,''
  JHEP {\bf 0509}, 070 (2005)
  [hep-th/0502186].




\bibitem{TailoringIV} 
  N.~Gromov and P.~Vieira,
  ``Tailoring Three-Point Functions and Integrability IV. Theta-morphism,''
  arXiv:1205.5288 [hep-th].




\bibitem{Kostovmorph} 
  Y.~Jiang, I.~Kostov, F.~Loebbert and D.~Serban,
  ``Fixing the Quantum Three-Point Function,''
  arXiv:1401.0384 [hep-th].




\bibitem{Pedrosl2}
P.~Vieira and T.~Wang,
  ``Tailoring Non-Compact Spin Chains,''
  [hep-th/1311.6404].
  


\bibitem{Kazakov3ptsl2} 
  V.~Kazakov and E.~Sobko,
 ``Three-point correlators of twist-2 operators in N=4 SYM at Born approximation,''
  JHEP {\bf 1306}, 061 (2013)
  [arXiv:1212.6563 [hep-th]].



\bibitem{FodaNovo} 
  O.~Foda, Y.~Jiang, I.~Kostov and D.~Serban,
  ``A tree-level 3-point function in the $su(3)$-sector of planar $N=4$ SYM,''
  arXiv:1302.3539 [hep-th].




\bibitem{StaudacherTwoloop} 
  M.~Staudacher,
  ``The Factorized S-matrix of CFT/AdS,''
  JHEP {\bf 0505}, 054 (2005)
  [hep-th/0412188].



\bibitem{Sotkov1} 
  G.~M.~Sotkov and R.~P.~Zaikov,
  ``Conformal Invariant Two Point and Three Point Functions for Fields with Arbitrary Spin,''
  Rept.\ Math.\ Phys.\  {\bf 12}, 375 (1977).
  


\bibitem{Sotkov2} 
  G.~M.~Sotkov and R.~P.~Zaikov,
  ``On the Structure of the Conformal Covariant $N$ Point Functions,''
  Rept.\ Math.\ Phys.\  {\bf 19}, 335 (1984).
  


\bibitem{Spinning} 
  M.~S.~Costa, J.~Penedones, D.~Poland and S.~Rychkov,
  ``Spinning Conformal Correlators,''
  JHEP {\bf 1111}, 071 (2011)
  [arXiv:1107.3554 [hep-th]].




\bibitem{BeisertTwoloop} 
  N.~Beisert,
  ``The su(2|3) dynamic spin chain,''
  Nucl.\ Phys.\ B {\bf 682}, 487 (2004)
  [hep-th/0310252].



\bibitem{Zarembo} 
  J.~A.~Minahan and K.~Zarembo,
  ``The Bethe ansatz for N=4 superYang-Mills,''
  JHEP {\bf 0303}, 013 (2003)
  [hep-th/0212208].




\bibitem{Minwalla} 
  S.~Lee, S.~Minwalla, M.~Rangamani and N.~Seiberg,
  ``Three point functions of chiral operators in D = 4, N=4 SYM at large N,''
  Adv.\ Theor.\ Math.\ Phys.\  {\bf 2}, 697 (1998)
  [hep-th/9806074].
  



\bibitem{Wheelersu3}
M.~Wheeler,  
``Scalar products in generalized models with SU(3)-symmetry,"
[arXiv:1204.2089  [math-ph]].




\bibitem{WheelerNovo} 
  M.~Wheeler,
  ``Multiple integral formulae for the scalar product of on-shell and off-shell Bethe vectors in SU(3)-invariant models,''
  Nucl.\ Phys.\ B {\bf 875}, 186 (2013)
  [arXiv:1306.0552 [math-ph]].




\bibitem{Slavnov1}
S.~Pakuliak,  E.~Ragoucy  and N.~A.~Slavnov,
``Scalar products in models with GL(3) trigonometric R-matrix. Highest coefficient,"
[arXiv:1311.3500  [math-ph]].



\bibitem{Slavnov2}
S.~Belliard, S.~Pakuliak,  E.~Ragoucy  and N.~A.~Slavnov,
``Highest coefficient of scalar products in SU(3)-invariant integrable models,"
J. Stat. Mech. (2012) P09003,
[arXiv:1206.4931  [math-ph]].



\bibitem{formfactors1} 
  T.~Klose and T.~McLoughlin,
  ``Worldsheet Form Factors in AdS/CFT,''
  Phys.\ Rev.\ D {\bf 87}, 026004 (2013)
  [arXiv:1208.2020 [hep-th]].



\bibitem{formfactors2} 
  T.~Klose and T.~McLoughlin,
  ``Comments on World-Sheet Form Factors in AdS/CFT,''
  arXiv:1307.3506.



\bibitem{fluxtube1} 
  B.~Basso, A.~Sever and P.~Vieira,
  ``Space-time S-matrix and Flux-tube S-matrix at Finite Coupling,''
  Phys.\ Rev.\ Lett.\  {\bf 111}, 091602 (2013)
  [arXiv:1303.1396 [hep-th]].




\bibitem{fluxtube2} 
  B.~Basso, A.~Sever and P.~Vieira,
  ``Space-time S-matrix and Flux tube S-matrix II. Extracting and Matching Data,''
  JHEP {\bf 1401}, 008 (2014)
  [arXiv:1306.2058 [hep-th]].



\bibitem{fluxtube3} 
  B.~Basso, A.~Sever and P.~Vieira,
  ``Space-time S-matrix and Flux-tube S-matrix III. The two-particle contributions,''
  arXiv:1402.3307 [hep-th].



\bibitem{Georgiou:2008vk}
  G.~Georgiou, V.~L.~Gili and R.~Russo,
  ``Operator Mixing and the AdS/CFT correspondence,''
  JHEP {\bf 0901} (2009) 082
  [arXiv:0810.0499 [hep-th]].



\bibitem{Georgiou:2012zj} 
  G.~Georgiou, V.~Gili, A.~Grossardt and J.~Plefka,
  ``Three-point functions in planar N=4 super Yang-Mills Theory for scalar operators up to length five at the one-loop order,''
  JHEP {\bf 1204}, 038 (2012)
  [arXiv:1201.0992 [hep-th]].




\bibitem{Korchemsky}
  B.~Eden, P.~Heslop, G.~P.~Korchemsky and E.~Sokatchev,
  Nucl.\ Phys.\ B {\bf 869} (2013) 329
  [arXiv:1103.3714 [hep-th]].



\bibitem{BeisertBMN} 
  N.~Beisert, C.~Kristjansen, J.~Plefka, G.~W.~Semenoff and M.~Staudacher,
  ``BMN correlators and operator mixing in N=4 superYang-Mills theory,''
  Nucl.\ Phys.\ B {\bf 650}, 125 (2003)
  [hep-th/0208178].




\bibitem{Georgiou:2004ty} 
  G.~Georgiou and G.~Travaglini,
  ``Fermion BMN operators, the dilatation operator of N=4 SYM, and pp wave string interactions,''
  JHEP {\bf 0404}, 001 (2004)
  [hep-th/0403188].




\bibitem{BeisertThesis} 
  N.~Beisert,
  ``The Dilatation operator of N=4 super Yang-Mills theory and integrability,''
  Phys.\ Rept.\  {\bf 405}, 1 (2004)
  [hep-th/0407277].



\end{thebibliography}
\end{document}